\documentclass[
reprint,     % Double column if uncommented
superscriptaddress,
showkeys,
amsmath,amssymb,
aps,
pra,
floatfix,
]{revtex4-2}

\usepackage{natbib}
\usepackage{graphicx}
\usepackage{dcolumn}
\usepackage{bm}
\usepackage{svg}
\usepackage[hidelinks]{hyperref}
\usepackage[mathlines]{lineno}
\usepackage{multirow}
\usepackage{subcaption}

\usepackage{lineno}
\usepackage{comment}
\usepackage{color}

\usepackage{cuted}
\usepackage{here}
\usepackage{booktabs}
\usepackage[subrefformat=parens]{subcaption}
\usepackage[justification=raggedright,singlelinecheck=false]{caption}
\usepackage[inline,final]{showlabels}

\showlabels{all}
\usepackage{soul}
\setstcolor{red}

\newcommand{\period}{\,\mathrm{.}}
\newcommand{\comma}{\,\mathrm{,}}
\begin{document}

\title{Quantum Noise from Vacuum Field Injection in Optical Cavities with Diffraction-related Loss}

\author{Kurumi Umemura}
\thanks{These authors contributed equally to this work.}
\author{Tomohiro Ishikawa}
\thanks{These authors contributed equally to this work.}
\author{Kenji Tsuji}
\thanks{These authors contributed equally to this work.}

\author{Shoki Iwaguchi}
\affiliation{Department of Physics, Nagoya University, Furo-cho, Chikusa-ku, Nagoya, Aichi 464-8602, Japan}

\author{Yutaro Enomoto}
\affiliation{Institute of Space and Astronautical Science, Japan Aerospace Exploration Agency, Sagamihara, Kanagawa 252-5210, Japan}

\author{Yuta Michimura}
\affiliation{Research Center for the Early Universe (RESCEU), School of Science, University of Tokyo, Tokyo 113-0033, Japan}
\affiliation{Kavli Institute for the Physics and Mathematics of the Universe (Kavli IPMU), WPI, UTIAS, University of Tokyo, Kashiwa, Chiba 277-8568, Japan}

\author{Kentaro Komori}
\affiliation{Research Center for the Early Universe (RESCEU), School of Science, University of Tokyo, Tokyo 113-0033, Japan}

\author{Keiko Kokeyama}
\affiliation{Department of Physics, Nagoya University, Furo-cho, Chikusa-ku, Nagoya, Aichi 464-8602, Japan}
\affiliation{The Kobayashi-Maskawa Institute for the Origin of Particles and the Universe, Nagoya University, Nagoya, Aichi 464-8602, Japan}
\affiliation{Cardiff University, Main Building, Park Place, Cardiff CF10 3AT, Wales, United Kingdom}

\author{Seiji Kawamura}
\affiliation{Department of Physics, Nagoya University, Furo-cho, Chikusa-ku, Nagoya, Aichi 464-8602, Japan}

\begin{abstract}
The space-based gravitational wave detector DECIGO is designed to observe primordial gravitational waves with $1,000$ km Fabry-Perot cavities. Its sensitivity is limited by quantum noise, and although squeezing can suppress it, its effectiveness is reduced by diffraction-related loss, which leads to the injection of vacuum fields into the interferometer. This paper presents a rigorous treatment of quantum field propagation in the presence of diffraction and higher-order mode losses, deriving input-output relations, and modeling their impact via an optomechanical block diagram. The analysis shows that diffraction-induced vacuum fields slightly increase radiation pressure noise, while shot noise remains unaffected. Nevertheless, cavity detuning with homodyne detection yields a dip in the noise spectrum. By accurately capturing these effects, this framework enables a detailed study of sensitivity improvements made by either just detuning the main cavity while implementing homodyne detection, or by combining this with optical-spring quantum locking using auxiliary cavities, laying a firm foundation for enhancing DECIGO’s capability to detect primordial gravitational waves.
\end{abstract}

\maketitle

%%%%%%%%%%%%%%%%%%%%%%%%%%  body  %%%%%%%%%%%%%%%%%%%%%%%%%%

%%=%-%=%=%=%=%=%=%=%=%=%=%=%-%=%=%=%=%=%=%=%=%=%=%=%-%=%=%=%=%=%=%=%=%=%=%=%-%=%=%=%=%=%=%=%=%=%=%=%-%=%=%=%=%=%=%=%=%=%=%=%-%=%=%=%=%=%=%=%=%=%=%
%%=%-%=%=%=%=%=%=%=%=%=%=%=%-%=%=%=%=%=%=%=%=%=%=%=%-%=%=%=%=%=%=%=%=%=%=%=%-%=%=%=%=%=%=%=%=%=%=%=%-%=%=%=%=%=%=%=%=%=%=%=%-%=%=%=%=%=%=%=%=%=%=%
%%=%-%=%=%=%=%=%=%=%=%=%=%=%-%=%=%=%=%=%=%=%=%=%=%=%-%=%=%=%=%=%=%=%=%=%=%=%-%=%=%=%=%=%=%=%=%=%=%=%-%=%=%=%=%=%=%=%=%=%=%=%-%=%=%=%=%=%=%=%=%=%=%
\section{Introduction}\label{Introduction}
The DECi-hertz Interferometer Gravitational wave Observatory (DECIGO) is a space interferometer with $1,000$ km Fabry-Perot arm cavities, and its frequency band is between 0.1 and 10 Hz \cite{PhysRevLett.87.221103,10.1093/ptep/ptab019,galaxies12020013}. The current conceptual design and the target sensitivity of DECIGO were established so as to achieve its primary science target of detecting primordial gravitational waves (PGWs), facilitating the revelation of many details concerning cosmic inflation. However, recent observations of the cosmic microwave background (CMB) by the Planck satellite have lowered the upper limit of PGWs \cite{akrami2020planck}, making it necessary to revisit the target sensitivity of DECIGO.\par
Since the default sensitivity of DECIGO is limited by quantum noise (radiation pressure noise and shot noise) in its frequency band, a seemingly straightforward remedy is to implement squeezing techniques into DECIGO’s interferometer design. However, the effectiveness of the standard squeezing technique, such as the cancelling of radiation pressure noise by utilizing ponderomotive squeezing and cavity detuning with homodyne detection, was expected to be less effective. This is because of the injection of vacuum fields into the squeezed state, which is caused by the significant diffraction-related loss of the laser field in the 1000 km-long arm cavity. Therefore, in previous research, simple parameter optimizations \cite{galaxies9010009,galaxies9010014,galaxies10010025} and optical-spring quantum locking \cite{PhysRevLett.90.083601,Antoine_Heidmann_2004,YAMADA2020126626,YAMADA2021127365,galaxies11060111,PhysRevD.107.022007,Ishikawa_2024} have been considered to improve DECIGO’s target sensitivity instead of using the conventional squeezing technique. Here, optical-spring quantum locking is a technique in which the radiation pressure noise is canceled at a specific frequency by attaching short, detuned auxiliary cavities (thus, not contaminated by the vacuum field caused by diffraction-related loss) to both mirrors of DECIGO’s arm cavity.\par
Nonetheless, it is important to establish a rigorous analysis of the contamination of the squeezed field by the vacuum field caused by diffraction-related loss to evaluate the effectiveness of reducing quantum noise through ponderomotive squeezing and cavity detuning, with homodyne detection in the main cavity. This analysis is also expected to provide a framework in which we can estimate how integrating cavity detuning with homodyne detection for the main cavity with optical-spring quantum locking using auxiliary cavities will further improve the expected sensitivity of DECIGO.\

Quantum noise in gravitational-wave interferometers is commonly analyzed using the two-photon formalism introduced by C.~M.~Caves \cite{PhysRevA.31.3068} and further developed in later studies \cite{PhysRevD.65.022002,PhysRevD.64.042006,PhysRevA.72.013818}. In this framework, the mixing of vacuum fields associated with optical losses is already incorporated in a systematic way. In most previous research, however, optical loss is treated as an effective mirror loss characterized by a single round-trip loss parameter. In this work, we extend this framework by explicitly modeling vacuum-field mixing associated with diffraction-related losses which are expected to be significant for extremely long-baseline cavities such as those of DECIGO.\par

In this paper, we define diffraction-related losses to include both “diffraction loss” and “higher-order mode loss”. Diffraction loss occurs when the mirror is smaller than the laser beam and part of the beam is clipped at the edges, while higher-order mode loss is caused by diffraction-clipped beams not fully converting back into the cavity’s resonant mode. Diffraction loss can occur even with a single mirror, whereas higher-order mode loss arises only in the presence of a cavity. We treat these losses separately to clarify their different physical origins. We also consider standard “mirror optical loss,” which originates from absorption or scattering at the mirror surface, as a reference case for the treatment of diffraction-related losses.\par
Further, we present the input-output relations including contributions from vacuum fields due to these losses, and evaluate the resulting quantum noise for DECIGO using ponderomotive squeezing and cavity detuning with homodyne detection.  This is the first rigorous study to consider the mixing of vacuum fields in a cavity with diffraction-related losses. Although this study is motivated by DECIGO, the framework developed here is broadly applicable to any optical cavity in which diffraction-related losses affect quantum noise, because it captures the essential physics of vacuum field injection due to such losses.\par
Section~\ref{sec:VF} introduces the vacuum field corresponding to three types of losses: mirror optical loss, diffraction loss, and higher-order mode loss. Section~\ref{sec:3} presents the block diagram and simulation of the above-mentioned relations. Sections~\ref{sec:discussion} and~\ref{sec:summary} present the discussion and conclusion respectively.

\begin{comment}
\begin{figure}[t]
    \begin{center}
    \includegraphics[width=0.5\linewidth]{fig/figDECIGO.JPG}
    \caption{Preliminary conceptual design of DECIGO \cite{10.1093/ptep/ptab019}. }
    \label{fig.DECIGO}
    \end{center}
\end{figure}
\end{comment}

%%=%-%=%=%=%=%=%=%=%=%=%=%=%-%=%=%=%=%=%=%=%=%=%=%=%-%=%=%=%=%=%=%=%=%=%=%=%-%=%=%=%=%=%=%=%=%=%=%=%-%=%=%=%=%=%=%=%=%=%=%=%-%=%=%=%=%=%=%=%=%=%=%
%%=%-%=%=%=%=%=%=%=%=%=%=%=%-%=%=%=%=%=%=%=%=%=%=%=%-%=%=%=%=%=%=%=%=%=%=%=%-%=%=%=%=%=%=%=%=%=%=%=%-%=%=%=%=%=%=%=%=%=%=%=%-%=%=%=%=%=%=%=%=%=%=%
%%=%-%=%=%=%=%=%=%=%=%=%=%=%-%=%=%=%=%=%=%=%=%=%=%=%-%=%=%=%=%=%=%=%=%=%=%=%-%=%=%=%=%=%=%=%=%=%=%=%-%=%=%=%=%=%=%=%=%=%=%=%-%=%=%=%=%=%=%=%=%=%=%
\section{The method of mixing vacuum fields}\label{sec:VF}
To understand how quantum noise manifests in the sensitivity curve of DECIGO, it is essential to analyze how vacuum fields are coupled into the interferometer through different kinds of optical loss. In this section, we introduce a general formalism for quantum fluctuations in Section~\ref{subsec:QF}, followed by a detailed discussion of three types of optical loss: mirror optical loss in Section~\ref{subsec:LoM}, diffraction loss in Section~\ref{subsec:LoL}, and the combined effect of diffraction loss and higher-order mode loss in the context of the entire cavity in Section~\ref{subsec:CwDL} since the two losses always occur with each other, whereas mirror optical loss is omitted for simplicity.

%%--#--#--#--#--#--#--#--#--#--#--#--#--#--#--#--#--#--#--#--#--#--#--#--#--#--#--#--#--#--#--#--#--#--#--#--#--#--#--#--#--#--#--#--#--#--#--#--%
%%--#--#--#--#--#--#--#--#--#--#--#--#--#--#--#--#--#--#--#--#--#--#--#--#--#--#--#--#--#--#--#--#--#--#--#--#--#--#--#--#--#--#--#--#--#--#--#--%
\subsection{Quantum fluctuations}\label{subsec:QF}
According to Ref.~\cite{schleich2011quantum}, quantum fluctuations can be described using quadrature-phase amplitudes, which separate the fluctuations into the phase quadrature $\hat{p}$ and the amplitude quadrature $\hat{q}$. Given the annihilation and creation operators of the quantum field, $\hat{a}(\omega)$ and ${\hat{a}}^{\dagger}(\omega)$, the single-photon quadratures $p^{\mathrm{(s)}}$ and $q^{\mathrm{(s)}}$ satisfy the following relations:
\begin{align}
\label{eq:AQ}
  &\hat{a}(\omega) = \hat{q}^{\mathrm{(s)}}+i\hat{p}^{\mathrm{(s)}}\\
  \label{eq:PQ}
  &\hat{q}^{\mathrm{(s)}} = {\frac{1}{2}}\Bigl(\hat{a}(\omega)+{\hat{a}}^{\dagger}(\omega)\Bigr),\hspace{3mm}
  \hat{p}^{\mathrm{(s)}} = {\frac{1}{2i}}\Bigl(\hat{a}(\omega)-{\hat{a}}^{\dagger}(\omega)\Bigr)\period
\end{align}
Here, $\omega$ corresponds to the sideband frequency, and the operators $\hat{a}(\omega)$ and ${\hat{a}}^{\dagger}(\omega)$ satisfy the following commutation relation:
\begin{align}
  \label{eq:CR1}
  \Bigl[\hat{a}(\omega),\ {\hat{a}}^{\dagger}({\omega}^{\prime})\Bigr]&={\delta}(\omega-{\omega}^{\prime}),\hspace{3mm} \Bigl[\hat{a}(\omega),\ {\hat{a^{\prime}}}^{\dagger}({\omega})\Bigr]=0\period
\end{align}
Note that since optical loss affects both quadratures equally, the following discussion focuses on the operator of the entire quantum field, ${\hat{a}}$. For clarity, quantum fields at different locations will be denoted by different symbols, such as $\hat{\alpha}$, $\hat{\beta}$, and so on, instead of using $\hat{a}$.
%%--#--#--#--#--#--#--#--#--#--#--#--#--#--#--#--#--#--#--#--#--#--#--#--#--#--#--#--#--#--#--#--#--#--#--#--#--#--#--#--#--#--#--#--#--#--#--#--%
%%--#--#--#--#--#--#--#--#--#--#--#--#--#--#--#--#--#--#--#--#--#--#--#--#--#--#--#--#--#--#--#--#--#--#--#--#--#--#--#--#--#--#--#--#--#--#--#--%
\begin{figure}[t]
  \centering
  \begin{minipage}[t]{\columnwidth}
    \centering
    \includegraphics[height=50mm]{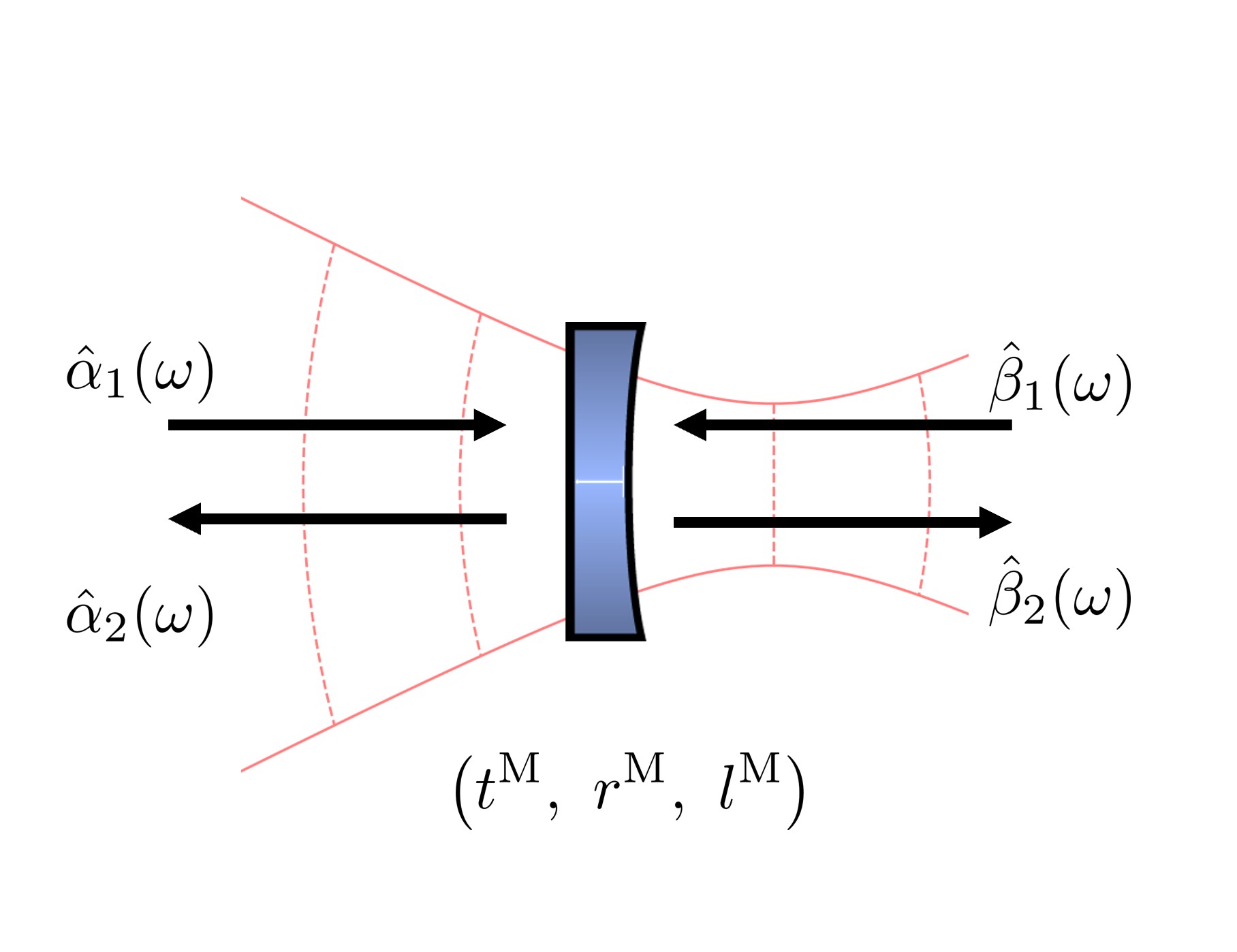}
    \subcaption{Mirror optical loss case.}
    \label{fig:w/o_DL}
  \end{minipage}

  \begin{minipage}[t]{\columnwidth}
    \centering
    \includegraphics[height=50mm]{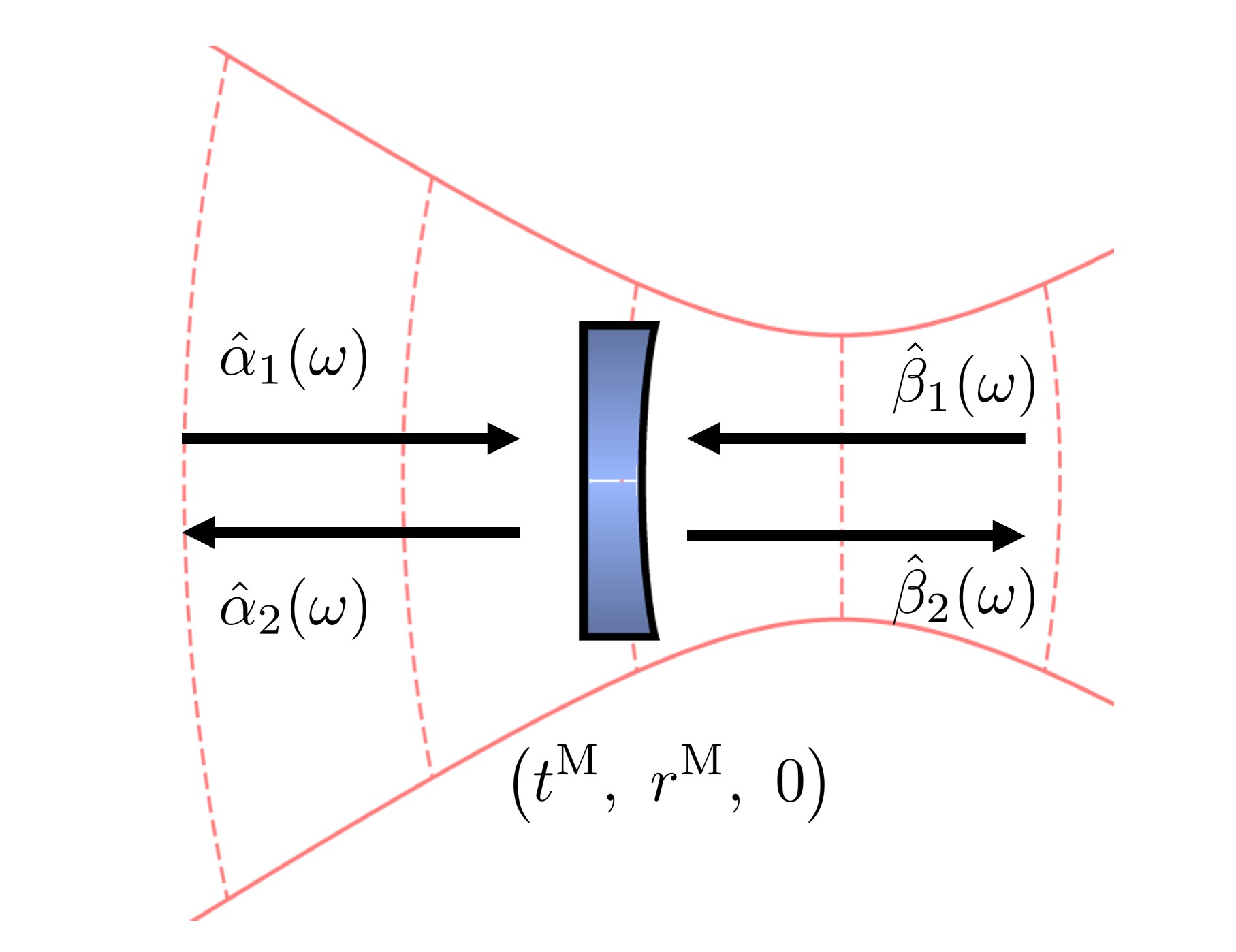}
    \subcaption{Diffraction loss case.}
    \label{fig:w/_DL}
  \end{minipage}

  \caption{Diagram of two sets of input and output operators of a mirror. Figure \ref{fig:w/o_DL} illustrates a mirror with optical loss, whose amplitude transmissivity, amplitude reflectivity, and optical loss are denoted by $t^{\rm{M}}({\omega})$, $r^{\rm{M}}({\omega})$, and $l^{\rm{M}}({\omega})$, respectively. Figure \ref{fig:w/_DL} represents a situation in which the beam diameter is larger than the mirror size, causing diffraction loss. In this case, the mirror optical loss is set to $l^{\rm{M}}({\omega})=0$.}
  \label{fig.amirror}
\end{figure}

\subsection{Mirror optical loss}\label{subsec:LoM}
First, we consider mirror optical loss. In this context, the frequency-dependent amplitude transmissivity $t^{\rm{M}}({\omega})$, amplitude reflectivity $r^{\rm{M}}({\omega})$, and optical loss $l^{\rm{M}}({\omega})$ of the mirror satisfy the following relations:
\begin{equation}
  \label{eq.2-1}
  \left|t^{\rm{M}}({\omega})\right|^2 +\left|r^{\rm{M}}({\omega})\right|^2+\left|l^{\rm{M}}({\omega})\right|^2 = 1,
\end{equation}
where the subscript M indicates that we are focusing on the optical loss associated with a mirror. Under this condition, we consider the quantum fields on both sides of the mirror, as illustrated in Figure~\ref{fig:w/o_DL}. The quantum fields incident on the mirror from each side are denoted by $\hat{\alpha}_1({\omega})$ and $\hat{\beta}_1({\omega})$, while the quantum fields outgoing from the mirror are denoted by $\hat{\alpha}_2({\omega})$ and $\hat{\beta}_2({\omega})$, respectively. Using the formalism shown in Ref.~\cite{PhysRevA.57.2134} these input operators satisfy the following relation:
\begin{equation}
  \label{eq.2-3}
  \left[\hat{\alpha}_1(\omega),\hat{\beta}_1^{\dagger}({\omega}^{\prime})\right]= 0 =  \left[\hat{\beta}_1(\omega),\hat{\alpha}_1^{\dagger}({\omega}^{\prime})\right].
\end{equation}
In addition, the normalized quantum fields introduced due to the mirror optical loss are distinguished by subscripts L and R corresponding to the positions where they are injected, and are defined as $\hat{\psi}^{\rm{M}}_{\rm{L}}({\omega})$ and $\hat{\psi}^{\rm{M}}_{\rm{R}}({\omega})$. Using an arbitrary coefficient ${\kappa}^{\mathrm{M}}$, the output operators, $\hat{\alpha}_2({\omega})$ and $\hat{\beta}_2({\omega})$ are given by:
\begin{alignat}{4}
  \label{eq:alpha_M_out}
 &\hat{\alpha}_2({\omega}) &&={}& -r^{\rm{M}}({\omega})\hat{\alpha}_1({\omega})&+t^{\rm{M}}({\omega})\hat{\beta}_1({\omega})&&+{\kappa}_{\mathrm{L}}^{\mathrm{M}}\hat{\psi}^{\rm{M}}_{\rm{L}}({\omega})\ ,\\
  \label{eq:beta_M_out}
 &\hat{\beta}_2({\omega}) &&={}& t^{\rm{M}}({\omega})\hat{\alpha}_1({\omega})&+r^{\rm{M}}({\omega})\hat{\beta}_1({\omega})&&+{\kappa}_{\mathrm{R}}^{\mathrm{M}}\hat{\psi}^{\rm{M}}_{\rm{R}}({\omega})\period
\end{alignat}
Since input fields and vacuum fields are independent, the commutation relations of noise operators are given by:
\begin{equation}
  \begin{split}
  \label{eq.2-6}
  {}\left[\hat{\psi}^{\rm{M}}_{\rm{L}}(\omega),\hat{\psi}_{\rm{L}}^{\rm{M\dagger}}(\omega')\right]&={\frac{1-\left|t^{\rm{M}}(\omega)\right|^2-\left|r^{\rm{M}}(\omega)\right|^2}{\left|{\kappa}_{\mathrm{L}}^{\mathrm{M}}(\omega)\right|^2}}\delta(\omega-\omega')\\ &= \left|{\frac{l^{\rm{M}}(\omega)}{{\kappa}_{\mathrm{L}}^{\mathrm{M}}(\omega)}}\right|^2\delta(\omega-\omega')\\
  \end{split}
\end{equation}
Considering that the normalized quantum fields satisfy the commutation relation given in Equation \ref{eq:CR1}, we obtain, ${\kappa}_{\mathrm{L}}^{\mathrm{M}}(\omega)=l^{\rm{M}}(\omega)$. Similarly, taking symmetry into account, we also find ${\kappa}_{\mathrm{R}}^{\mathrm{M}}(\omega)=l^{\rm{M}}(\omega)$. Thus, rewriting Equations \ref{eq:alpha_M_out} and \ref{eq:beta_M_out} with ${\kappa}^{\mathrm{M}}(\omega)=l^{\rm{M}}(\omega)$, we obtain the following:
\begin{alignat}{4}
  \label{eq:alpha_M_out2}
 &\hat{\alpha}_2({\omega}) &&={}& -r^{\rm{M}}({\omega})\hat{\alpha}_1({\omega})&+t^{\rm{M}}({\omega})\hat{\beta}_1({\omega})&&+l^{\rm{M}}(\omega)\hat{\psi}^{\rm{M}}_{\rm{L}}({\omega})\ ,\\
  \label{eq:beta_M_out2}
 &\hat{\beta}_2({\omega}) &&={}& t^{\rm{M}}({\omega})\hat{\alpha}_1({\omega})&+r^{\rm{M}}({\omega})\hat{\beta}_1({\omega})&&+l^{\rm{M}}(\omega)\hat{\psi}^{\rm{M}}_{\rm{R}}({\omega})\period
\end{alignat}
As shown in Equations \ref{eq:alpha_M_out2} and \ref{eq:beta_M_out2}, the output operators are mixed with vacuum fields in addition to the input operators, with the mixing strength determined by the mirror optical loss, $l^{\rm{M}}$.

%%--#--#--#--#--#--#--#--#--#--#--#--#--#--#--#--#--#--#--#--#--#--#--#--#--#--#--#--#--#--#--#--#--#--#--#--#--#--#--#--#--#--#--#--#--#--#--#--%
%%--#--#--#--#--#--#--#--#--#--#--#--#--#--#--#--#--#--#--#--#--#--#--#--#--#--#--#--#--#--#--#--#--#--#--#--#--#--#--#--#--#--#--#--#--#--#--#--%
\subsection{Diffraction loss}\label{subsec:LoL}
In practice, laser beams possess a finite divergence in the transverse direction. Consequently, when propagating over the long baseline of an interferometer such as DECIGO, the beam gradually expands, and a portion of its power may spill over the finite size of the mirrors, resulting in diffraction loss, as shown in Figure \ref{fig:w/_DL}. Assuming that the transverse mode of the laser beam incident on the mirror consists solely of the fundamental Hermite–Gaussian mode TEM$_{00}$, this diffraction effect can be characterized by a diffraction factor $D$, which is obtained by integrating the TEM$_{00}$ mode over the finite size of the mirror:
\begin{equation}
  \label{eq.2-0}
  D^2 \equiv1-\exp\left(-\frac{2\pi z_{\rm{R}}}{\lambda(l^2+z_{\rm{R}}^2)}R^2 \right),
\end{equation}
where, $z_{\rm{R}}$ is Rayleigh length, $\lambda$ is the wavelength of the laser beam, $l$ is the distance away from the beam waist, and
$R$ is the mirror radius. Using this factor, for instance, if the quantum field without diffraction loss is denoted by $\hat{\alpha}_1(\omega)$, then the actual quantum field interacting with the mirror, ${\hat{\alpha}}_1^{\prime}(\omega)$, is expressed as ${\hat{\alpha}}_1^{\prime}(\omega) = D{\hat{\alpha}}_1(\omega)$.
\begin{comment}
\begin{equation}
  \label{eq.2-2-1}
  \left|t({\omega})\right|^2 +\left|r({\omega})\right|^2 = 1.
\end{equation}
\end{comment}
For simplicity, we assume hereafter that the mirror has no optical loss and that its amplitude transmissivity and reflectivity satisfy $\left|t({\omega})\right|^2 +\left|r({\omega})\right|^2 = 1$. Under this condition, the output operators, $\hat{\alpha}_2({\omega})$ and $\hat{\beta}_2({\omega})$, are represented by the input operators with diffraction loss as:
\begin{alignat}{4}
  \label{eq.2-2-2}
 &\hat{\alpha}_2({\omega}) &&={}& - r({\omega})D\hat{\alpha}_1({\omega})&+t({\omega})D\hat{\beta}_1({\omega})&&+{\kappa}_{\mathrm{L}}^{\mathrm{diff}}({\omega})\hat{\psi}^{\rm{diff}}_{\rm{L}}({\omega})\ ,\\
  \label{eq.2-2-3}
 &\hat{\beta}_2({\omega}) &&={}& t({\omega})D\hat{\alpha}_1({\omega})&-r({\omega})D\hat{\beta}_1({\omega})&&+{\kappa}_{\mathrm{L}}^{\mathrm{diff}}({\omega})\hat{\psi}^{\rm{diff}}_{\rm{R}}({\omega})\period
\end{alignat}
Here, similar to mirror optical loss, ${\kappa}^{\mathrm{diff}}({\omega})$ is an arbitrary coefficient and $\hat{\psi}_{\rm{L}}^{\rm{diff}}$ and $\hat{\psi}^{\rm{diff}}_{\rm{R}}$ are vacuum fields corresponding to diffraction loss. In addition, the commutation relations of noise operators are given by:
\begin{equation}
  \begin{split}
  \label{eq.2-2-5}
  \left[\hat{\psi}^{\rm{diff}}_{\rm{L}}(\omega),\hat{\psi}^{\rm{diff\dagger}}_{\rm{L}}(\omega')\right]&={\frac{1-D^2}{\left|{\kappa}_{\mathrm{L}}^{\mathrm{diff}}(\omega)\right|^2}}\delta(\omega-\omega')= \left|{\frac{U}{{\kappa}_{\mathrm{L}}^{\mathrm{diff}}(\omega)}}\right|^2,
  \end{split}
\end{equation}
where $U^2 = 1-D^2$. $U$ is a coefficient representing the diffraction loss. The relationship between $U$ and $D$ is obtained from the overlap integral of the TEM$_{00}$ mode and the mirror diameter. Following a similar discussion as in the previous section, by replacing ${\kappa}^{\mathrm{diff}}(\omega)$ with $U$ in Equations~\ref{eq.2-2-2} and~\ref{eq.2-2-3}, we obtain the following:
\begin{alignat}{4}
  \label{eq.2-2-8}
 &\hat{\alpha}_2({\omega}) &&={}& -r({\omega})D\hat{\alpha}_1({\omega})&+t({\omega})D\hat{\beta}_1({\omega})&+U\hat{\psi}^{\rm{diff}}_{\rm{L}}({\omega})\ ,\\
  \label{eq.2-2-9}
 &\hat{\beta}_2({\omega}) &&={}& t({\omega})D\hat{\alpha}_1({\omega})&-r({\omega})D\hat{\beta}_1({\omega})&+U\hat{\psi}^{\rm{diff}}_{\rm{R}}({\omega})\period
\end{alignat}

From Equations~\ref{eq.2-2-8} and~\ref{eq.2-2-9}, it can be seen that, in addition to the input operators, the output operators are mixed with vacuum fields corresponding to the diffraction effects, $U$.
%%--#--#--#--#--#--#--#--#--#--#--#--#--#--#--#--#--#--#--#--#--#--#--#--#--#--#--#--#--#--#--#--#--#--#--#--#--#--#--#--#--#--#--#--#--#--#--#--%
%%--#--#--#--#--#--#--#--#--#--#--#--#--#--#--#--#--#--#--#--#--#--#--#--#--#--#--#--#--#--#--#--#--#--#--#--#--#--#--#--#--#--#--#--#--#--#--#--%

\subsection{Cavity with diffraction loss and higher-order mode loss}\label{subsec:CwDL}
 \begin{figure}[t]
    \begin{center}
    \includegraphics[height=55mm]{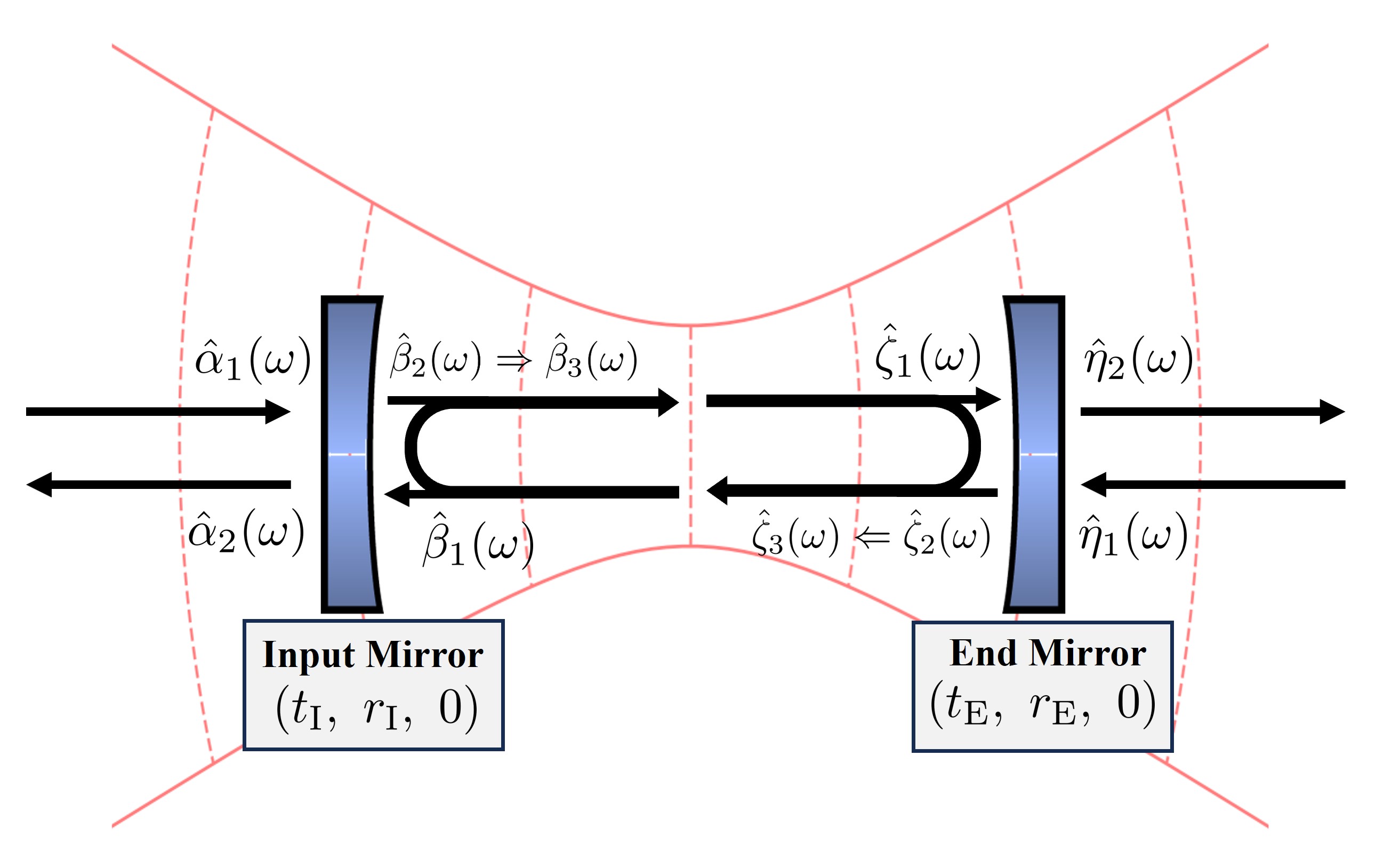}
    \caption{Diagram of two sets of input and output operators in a cavity with diffraction-related loss. Here, the two mirrors have no mirror optical loss, and only diffraction-related loss is considered. The quantum fields truncated due to diffraction-related loss, $\hat{\beta}_2({\omega})$ and $\hat{\zeta}_2({\omega})$, are reconstructed by the cavity eigenmodes and are represented as $\hat{\beta}_3({\omega})$ and $\hat{\zeta}_3({\omega})$, respectively.}
    \label{fig.3}
    \end{center}
\end{figure}
Finally, based on the content discussed in the previous sections, we define the quantum fields associated with a cavity. As shown in Figure \ref{fig.3}, the input mirror and end mirror that form the cavity are located at positions $x=-L/2$ and $x=L/2$. The quantum field outside the cavity on the input mirror side is denoted by $\hat{\alpha}({\omega})$, and the field on the end mirror side is denoted by $\hat{\eta}({\omega})$. The input annihilation and creation operators of the cavity satisfy the following commutation relations:
\begin{align}
  \label{eq.2-4-1}
  \left[\hat{\alpha}_1(\omega),\hat{\alpha}_1^{\dagger}(\omega')\right]=\delta(\omega-\omega') =  \left[\hat{\eta}_1(\omega),\hat{\eta}_1^{\dagger}(\omega')\right],\\
  \label{eq.2-4-2}
  \left[\hat{\alpha}_1(\omega),\hat{\eta}_1^{\dagger}({\omega}^{\prime})\right]= 0 =  \left[\hat{\eta}_1(\omega),\hat{\alpha}_1^{\dagger}({\omega}^{\prime})\right].
\end{align}
Here, the operators for the input mirror are represented by the input operators as follows:
\begin{alignat}{4}
  \label{eq:alpha2_cav}
 &\hat{\alpha}_2({\omega}) &&={}& -r_{\rm{I}}({\omega})D_{\rm{I}}\hat{\alpha}_1({\omega})&+t_{\rm{I}}({\omega})D_{\rm{I}}\hat{\beta}_1({\omega})&+U_{\rm{I}}\hat{\psi}^{\rm{diff}}_{\rm{IL}}({\omega})\comma\\
\label{eq:beta2_cav}
 &\hat{\beta}_2({\omega}) &&={}& t_{\rm{I}}({\omega})D_{\rm{I}}\hat{\alpha}_1({\omega})&+r_{\rm{I}}({\omega})D_{\rm{I}}\hat{\beta}_1({\omega})&+U_{\rm{I}}\hat{\psi}^{\rm{diff}}_{\rm{IR}}({\omega})\period
\end{alignat}
Similarly, the operators associated with the end mirror are given by:
\begin{alignat}{4}
  \label{eq:zeta2_cav}
  &\hat{\zeta}_2({\omega}) &&={}& t_{\rm{E}}({\omega})D_{\rm{E}}\hat{\eta}_1({\omega})&+r_{\rm{E}}({\omega})D_{\rm{E}}\hat{\zeta}_1({\omega})&&+U_{\rm{E}}\hat{\psi}^{\rm{diff}}_{\rm{EL}}({\omega})\comma\\
  \label{eq:eta2_cav}
  &\hat{\eta}_2({\omega}) &&={}& -r_{\rm{I}}({\omega})D_{\rm{E}}\hat{\eta}_1({\omega})&+t_{\rm{E}}({\omega})D_{\rm{E}}\hat{\zeta}_1({\omega})&&+U_{\rm{E}}\hat{\psi}^{\rm{diff}}_{\rm{ER}}({\omega})\period
\end{alignat}
Here, $\hat{\psi}^{\rm{diff}}_{\rm{IL}}({\omega})$, $\hat{\psi}^{\rm{diff}}_{\rm{IR}}({\omega})$, $\hat{\psi}^{\rm{diff}}_{\rm{EL}}({\omega})$ and  $\hat{\psi}^{\rm{diff}}_{\rm{ER}}({\omega})$
 are vacuum fields corresponding to diffraction-related loss, with L and R representing the direction of the movement. I and E represent the input and end mirrors, respectively. \par
Furthermore, optical cavity is typically designed to resonate only with the TEM$_{00}$ mode. If the incident light consists purely of the TEM$_{00}$, and the mirrors introduce significant diffraction loss, the reflected or transmitted beam at the mirrors may become an imperfect mode due to truncation at the spatial edges. In such cases, the TEM$_{00}$ component can be reconstructed by evaluating the overlap between the distorted spatial mode of the actual beam and the cavity eigenmode. While this allows recovery of the TEM$_{00}$ component defined by the cavity eigenmode, the remaining parts of the beam correspond to higher-order modes in this basis and are generally non-resonant. Therefore, if the cavity finesse is sufficiently high, these components can be regarded as an effective loss for the TEM$_{00}$ mode. This mechanism of mode conversion and subsequent vacuum field injection is a well-known issue even in ground-based gravitational wave detectors; for instance, it has been pointed out that scattering from mirror surface imperfections degrades the quantum state by coupling the fundamental mode into non-resonant higher-order modes \cite{PhysRevD.93.082004}. In our study, which specifically addresses the dominant geometric diffraction occurring in long-baseline cavities such as those of DECIGO, this corresponds to the coupling of vacuum fluctuations associated with higher-order mode loss, denoted as $\hat{\psi}^{\rm{HOM}}({\omega})$.\par

Here, if the truncated quantum field due to diffraction, as described in Section \ref{subsec:LoL}, is represented by $\hat{\beta}_2({\omega})$, then the reconstructed vacuum field $\hat{\beta}_3({\omega})$ is defined as follows:
\begin{equation}\label{eq:HOM}
  \hat{\beta}_3({\omega}) = D\hat{\beta}_2({\omega}) + {\kappa}^{\mathrm{HOM}}({\omega})\hat{\psi}^{\rm{HOM}}({\omega})\period
\end{equation}
Here, it should be noted that the parameter $D$, characterizing the reconstruction process, although derived through a different procedure, is equal to the value of $D$ presented in Section \ref{subsec:LoL}.\par
Moreover, considering that the input field and vacuum field are independent and that the output field undergoes a unitary transformation, the commutation relations of the noise operators, $\hat{\psi}^{\rm{HOM}}({\omega})$,  are given by:

\begin{equation}
  \begin{split}
  \label{eq:ex_HOM}
  \left[\hat{\psi}^{\rm{HOM}}(\omega),\hat{\psi}^{\rm{HOM\dagger}}({\omega}^{\prime})\right]&={\frac{1-D^2}{\left|{\kappa}^{\mathrm{HOM}}(\omega)\right|^2}}\delta(\omega-{\omega}^{\prime})\\&= \left|{\frac{U}{{\kappa}^{\mathrm{HOM}}(\omega)}}\right|^2\period
  \end{split}
\end{equation}
Therefore, the coefficient ${\kappa}^{\mathrm{HOM}}(\omega)$ of the vacuum field $\hat{\psi}^{\rm{HOM}}({\omega})$ corresponding to the higher-order mode loss is equal to the coefficient representing the diffraction-related loss $U$. This corresponds to the fact that the sum of vacuum fields converted into each mode equals $U$, and thus, within the range where $U$ is sufficiently small, the effect of quantum fields that have transitioned to higher-order modes returning back to the fundamental mode can be neglected. Since this study is based on this assumption, the vacuum field corresponding to the higher-order mode loss is uniquely described by Equation \ref{eq:HOM}, regardless of the number of reflections inside the cavity (one reflection, two reflections, etc.).  In this context, the quantum field $\hat{\beta}_3({\omega})$, which propagates inside the cavity after reconstruction and reaches the end mirror, undergoes only a simple phase change and is given by:
In this context, the quantum field $\hat{\beta}_3(\omega)$, which propagates inside the cavity after reconstruction and reaches the end mirror, undergoes only a simple phase change and is given by $\hat{\zeta}_1(\omega) = \hat{\beta}_3(\omega)e^{2i\theta(\omega)}$. Here, $\theta(\omega)$ is defined as $\theta(\omega) = \frac{\omega L}{2c}$ using the speed of light $c$. Similarly, the vacuum field $\hat{\zeta}_3(\omega)$ reconstructed at the end mirror propagates inside the cavity, and the quantum field $\hat{\beta}_1(\omega)$ upon reaching the input mirror satisfies the relation $\hat{\beta}_1(\omega) = \hat{\zeta}_3(\omega)e^{2i\theta(\omega)}$.

Based on Equations \ref{eq.2-4-1} through \ref{eq:ex_HOM}, the operators in the cavity satisfy the following commutation relations:

\begin{comment}
\begin{align}
\notag
\label{eq.2-4-12}
 \left[\hat{\beta}_1({\omega}),\hat{\beta}_1^\dagger({\omega}^{\prime})\right]
 &=\frac{1-\left|r_{\rm{I}}({\omega})D_{\rm{I}}^2r_{\rm{E}}({\omega})D_{\rm{E}}^2\right|^2}{\left|1-r_{\rm{I}}({\omega})D_{\rm{I}}^2r_{\rm{E}}({\omega})D_{\rm{E}}^2{\,e^{4i\theta({\omega})}}\right|^2}\delta(\omega-\omega')\\
 &= \left[\hat{\zeta}_1({\omega}),\hat{\zeta}_1^\dagger({\omega}')\right]\comma\\
\notag
\label{eq.2-4-13}
 \left[\hat{\beta}_1({\omega})e^{-i\theta},\hat{\zeta}_1^\dagger({\omega'})e^{i\theta}\right]
 &=\frac{r_{\rm{E}}({\omega})D_{\rm{E}}^2{\,e^{2i\theta({\omega})}}\left(1-\left|r_{\rm{I}}({\omega})D_{\rm{I}}^2\right|^2\right)+r_{\rm{I}}^\ast({\omega})D_{\rm{I}}^2{\,e^{-2i\theta({\omega})}}\left(1-\left|r_{\rm{E}}({\omega})D_{\rm{E}}^2\right|^2\right)}{\left|1-
 r_{\rm{I}}({\omega})D_{\rm{I}}^2r_{\rm{E}}({\omega})D_{\rm{E}}^2{\,e^{4i\theta({\omega})}}\right|^2}\delta(\omega-\omega')\\
 &= \left[\hat{\zeta}_1({\omega})e^{-i\theta},\hat{\beta}_1^\dagger({\omega}')e^{i\theta}\right]^\ast\period
\end{align}
\end{comment}

\begin{equation}
\begin{split}
\label{eq.2-4-12}
\left[\hat{\beta}_1({\omega}),\hat{\beta}_1^\dagger({\omega}^{\prime})\right]
&=\frac{1-\left|r_{\rm{I}}({\omega})D_{\rm{I}}^2r_{\rm{E}}({\omega})D_{\rm{E}}^2\right|^2}{S^2}\delta(\omega-\omega')\\
&= \left[\hat{\zeta}_1({\omega}),\hat{\zeta}_1^\dagger({\omega}')\right]\\
 S&= \left|1-
 r_{\rm{I}}({\omega})D_{\rm{I}}^2r_{\rm{E}}({\omega})D_{\rm{E}}^2{\,e^{4i\theta({\omega})}}\right|\comma
\end{split}
\end{equation}
\begin{equation}
\begin{split}
    \label{eq.2-4-13}
 &\left[\hat{\beta}_1({\omega})e^{-i\theta},\hat{\zeta}_1^\dagger({\omega'})e^{i\theta}\right]\\
 &=\left[\frac{r_{\rm{E}}({\omega})D_{\rm{E}}^2{\,e^{2i\theta({\omega})}}\left(1-\left|r_{\rm{I}}({\omega})D_{\rm{I}}^2\right|^2\right)}{S^2}\right.\\
 &+ \left.\frac{r_{\rm{I}}^\ast({\omega})D_{\rm{I}}^2{\,e^{-2i\theta({\omega})}}\left(1-\left|r_{\rm{E}}({\omega})D_{\rm{E}}^2\right|^2\right)}{S^2}\right]\delta(\omega-\omega')\\
 &= \left[\hat{\zeta}_1({\omega})e^{-i\theta},\hat{\beta}_1^\dagger({\omega}')e^{i\theta}\right]^\ast\\
\period
\end{split}
\end{equation}

Considering the effective amplitude reflectivity $r_{\rm{eff,i}} = r_{\rm{i}}D_{\rm{i}}^2$ and transmissivity $t_{\rm{eff,i}} = t_{\rm{i}}D_{\rm{i}}^2$, Equations \ref{eq.2-4-12} and \ref{eq.2-4-13} are consistent with the commutation relations of the cavity operators in the presence of optical loss, as given by Equation (9) in Ref.~\cite{PhysRevLett.77.1739}.

%%=%-%=%=%=%=%=%=%=%=%=%=%=%-%=%=%=%=%=%=%=%=%=%=%=%-%=%=%=%=%=%=%=%=%=%=%=%-%=%=%=%=%=%=%=%=%=%=%=%-%=%=%=%=%=%=%=%=%=%=%=%-%=%=%=%=%=%=%=%=%=%=%
%%=%-%=%=%=%=%=%=%=%=%=%=%=%-%=%=%=%=%=%=%=%=%=%=%=%-%=%=%=%=%=%=%=%=%=%=%=%-%=%=%=%=%=%=%=%=%=%=%=%-%=%=%=%=%=%=%=%=%=%=%=%-%=%=%=%=%=%=%=%=%=%=%
%%=%-%=%=%=%=%=%=%=%=%=%=%=%-%=%=%=%=%=%=%=%=%=%=%=%-%=%=%=%=%=%=%=%=%=%=%=%-%=%=%=%=%=%=%=%=%=%=%=%-%=%=%=%=%=%=%=%=%=%=%=%-%=%=%=%=%=%=%=%=%=%=%
\begin{figure*}[!t]
\centering

% --- top figure ---
\begin{minipage}{0.9\textwidth}
\centering
\includegraphics[height=105mm]{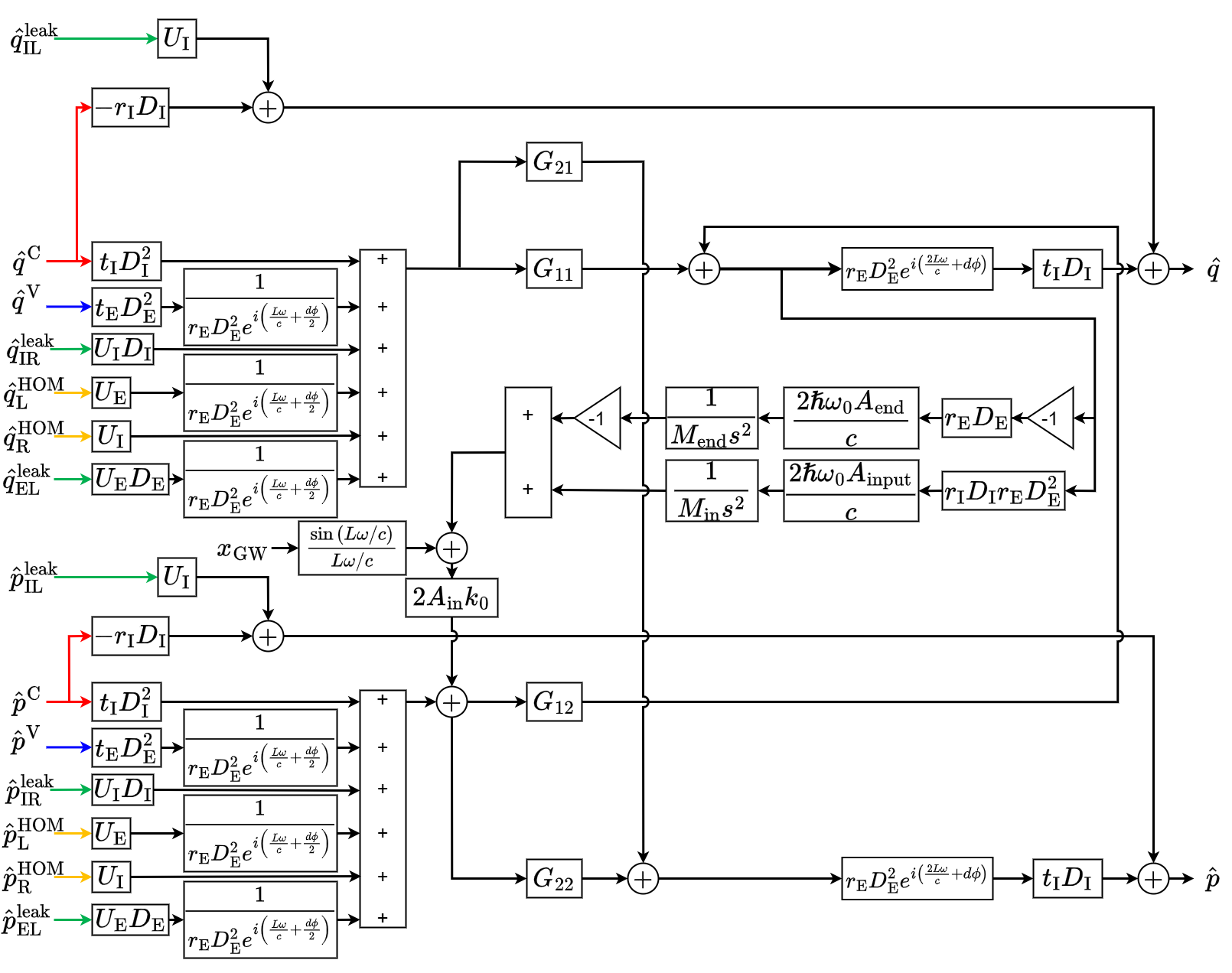}
\captionof{figure}{Block diagram of a DECIGO-like cavity with diffraction-related loss and an optical spring. The left side illustrates the quantum fluctuations injected into the cavity at various points. The subsequent blocks account for the amplitude and phase changes as they propagate to the detection point. The resulting quantum noise is represented by $p$ and $q$ on the right-hand side.}
\label{fig.4}
\end{minipage}

\vspace{5mm}

% --- bottom left ---
\begin{minipage}{0.42\textwidth}
\centering
\includegraphics[width=\linewidth]{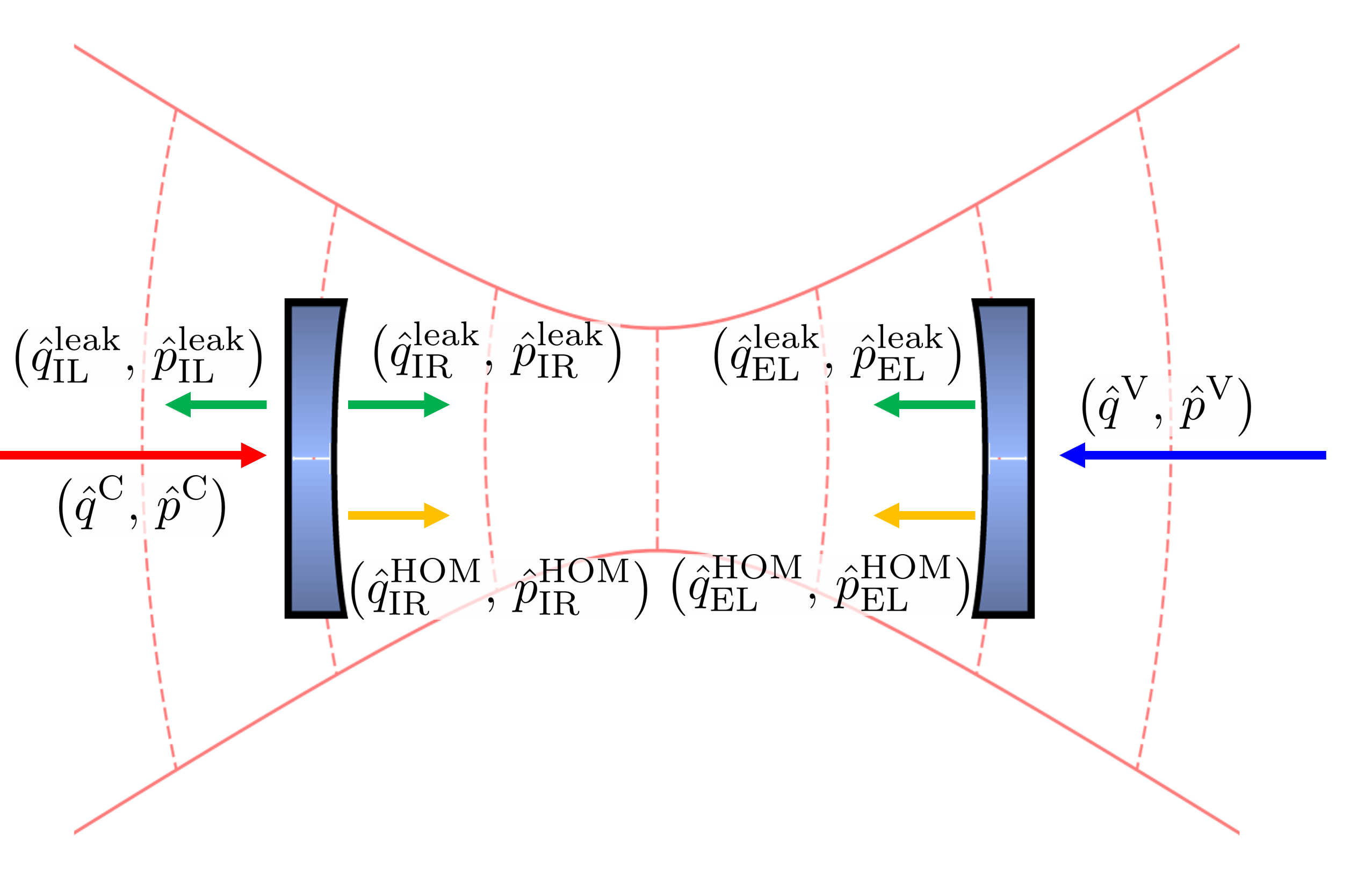}
\captionof{figure}{Optical configuration corresponding to the block diagram (Figure \ref{fig.4}).}
\label{fig:optical_scheme}
\end{minipage}
\hfill
% --- bottom right ---
\begin{minipage}{0.54\textwidth}
\captionof{table}{Quantum fields appearing in the model.}
\label{table:quantum_fields}
\centering
\begin{tabular}{lll}
\toprule
Symbol & Type & Origin \\
\midrule
${\hat{q}}^{\mathrm{C}},\, {\hat{p}}^{\mathrm{C}}$ & Laser & Input laser \\
${\hat{q}}^{\mathrm{V}},\, {\hat{p}}^{\mathrm{V}}$ & Empty port & Behind ETM \\
${\hat{q}}^{\mathrm{leak}}_{\mathrm{IL}},\, {\hat{p}}^{\mathrm{leak}}_{\mathrm{IL}}$ & Diffraction & Left side of ITM \\
${\hat{q}}^{\mathrm{leak}}_{\mathrm{IR}},\, {\hat{p}}^{\mathrm{leak}}_{\mathrm{IR}}$ & Diffraction & Right side of ITM \\
${\hat{q}}^{\mathrm{leak}}_{\mathrm{EL}},\, {\hat{p}}^{\mathrm{leak}}_{\mathrm{EL}}$ & Diffraction & Left side of ETM \\
${\hat{q}}^{\mathrm{HOM}}_{\mathrm{IR}},\, {\hat{p}}^{\mathrm{HOM}}_{\mathrm{IR}}$ & HOM coupling & Right side of ITM \\
${\hat{q}}^{\mathrm{HOM}}_{\mathrm{EL}},\, {\hat{p}}^{\mathrm{HOM}}_{\mathrm{EL}}$ & HOM coupling & Right side of ETM \\
\bottomrule
\end{tabular}
\end{minipage}
\end{figure*}

\begin{figure*}[!t]
    \begin{center}
    \includegraphics[height=80mm]{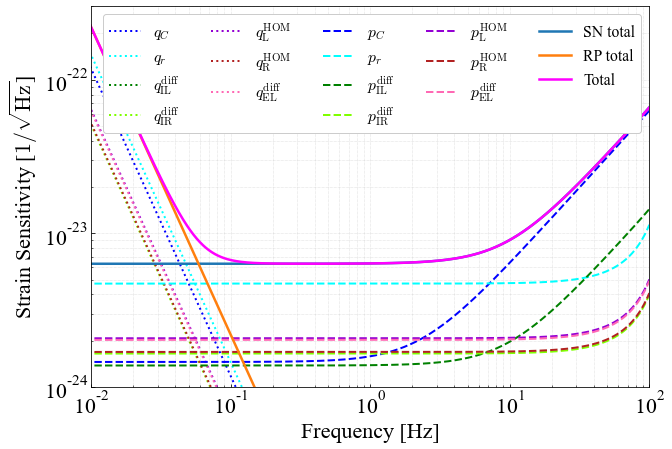}
    \caption{Spectra of quantum noise contributions from various sources, expressed in terms of strain sensitivity, for DECIGO, calculated from the block diagram (Figure~\ref{fig.4}). Dotted curves represent noise arising from various kinds of quantum fluctuations in the amplitude quadrature, with their total quadratic sum indicated by the orange solid curve (labeled “RP total”). Similarly, dashed curves represent noise from various kinds of quantum fluctuations in the phase quadrature, with their total quadratic sum shown as the blue solid curve (“SN total”). The resulting total quantum noise curve is depicted in magenta (“Total”).}
    \label{fig.5}
    \end{center}
\end{figure*}

\section{Propagation of quantum fields in a cavity}\label{sec:3}
In the previous section, we discussed the modification of quantum fields due to optical loss. In this section, we describe how quantum fields, including both the intrinsic fluctuations of the laser and those introduced through optical loss, propagate and contribute to noise through optomechanical interactions within the cavity.\par
First, even when the injection process is identical, the propagation of quantum fluctuations in a cavity must be described using quadratures with different phases and amplitudes. This is because amplitude fluctuations interact with the mirrors and induce fluctuations in the optical path length. These path-length fluctuations are then converted into phase fluctuations. As a result, the amplitude and phase components evolve differently. Moreover, when the cavity is detuned, phase fluctuations are also partially converted into amplitude fluctuations. This coupling forms a loop, and the quantum field emerging from the cavity becomes highly complex.
To systematically describe these effects, we employ the two-photon formalism \cite{PhysRevA.31.3068} as an effective response function. The quadrature operators in the two-photon formalism are defined as
\begin{align}
    \hat{q} &= {\frac{1}{\sqrt{2}}}\left[\hat{a}\left(\omega_0+\omega\right) + \hat{a}^\dagger\left(\omega_0-\omega\right)\right]\\
    \hat{p} &= {\frac{1}{i\sqrt{2}}}\left[\hat{a}\left(\omega_0+\omega\right) - \hat{a}^\dagger\left(\omega_0-\omega\right)\right]
\end{align}
from the single photon annihilation operators $\hat{a}\left(\omega_0{\pm}\omega\right)$ in Equation \ref{eq:AQ}. Using this formalism, the input–output relation of the quantum field in the cavity can be expressed as follows \cite{danilishin2012quantum,PhysRevD.88.022002}:
\begin{equation}
  \begin{split}
  \label{eq.3-2}
  \begin{bmatrix} {\hat{q}}_{\rm{out}} \\ {\hat{p}}_{\rm{out}} \end{bmatrix} &= V^{-1}
  \begin{bmatrix}  g^\ast(\omega)& 0 \\ 0 &g(-\omega) \end{bmatrix}V
  \begin{bmatrix}  {\hat{q}}_{\rm{in}}\\{\hat{p}}_{\rm{in}}  \end{bmatrix}= G(\omega)  \begin{bmatrix}  {\hat{q}}_{\rm{in}}\\{\hat{p}}_{\rm{in}}  \end{bmatrix}\\
  V &= \frac{1}{\sqrt{2}} \begin{bmatrix} 1 & 1 \\ -{\rm{i}} &{\rm{i}} \end{bmatrix}\\
G(\omega) &\equiv \frac{1}{2} \begin{bmatrix} g^\ast(\omega)+g(-\omega)& {\rm{i}}[g^\ast(\omega)-g(-\omega)] \\
  -{\rm{i}}[g^\ast(\omega)-g(-\omega)] & g^\ast(\omega)+g(-\omega) \end{bmatrix}\period
  \end{split}
\end{equation}
Here, $g(\omega)$ is given by:
\begin{equation}
  \label{eq:g_omega}
  g(\omega) = \frac{1}{1-r_{\rm{I}}({\omega})D_{\rm{I}}^2r_{\rm{E}}({\omega})D_{\rm{E}}^2\exp\left[-{\rm{i}}\left(d\phi+\frac{2L}{c}\omega\right)\right]}\comma
\end{equation}
where $d\phi$ denotes the detuning angle of the cavity. Note that Equation \ref{eq:g_omega} differs slightly from the corresponding expression in previous studies \cite{galaxies9010009,galaxies9010014,galaxies10010025}, as it explicitly depends on the diffraction-related coefficients $D$, because diffraction-related losses include both diffraction loss and higher-order mode loss.\par
Next, we calculate the propagation of quantum fields throughout the entire system using the block diagram shown in Figure \ref {fig.4}. Figure \ref{fig:optical_scheme} also shows the corresponding optical scheme. In addition, Table \ref{table:quantum_fields} summarizes the definitions of the quantum fields. Here, $q^{\rm j} _{\rm i}$ and $p^{\rm j} _{\rm i}$ represent independent quantum fluctuations, where the superscript indicates the origin and the subscript indicates the point of entry. The terms $q^{\mathrm{C}}$ and $p^{\mathrm{C}}$ correspond to the quantum fields associated with the carrier light, and $q^{\mathrm{V}}$ and $p^{\mathrm{V}}$ represent vacuum fields entering from the empty port on the end mirror. Assuming that each mirror has no mirror optical loss, meaning that $r_k^2+t_k^2=1$, the discussion in Section \ref{sec:VF} shows that vacuum fields enter at four points due to diffraction loss and at two points due to higher-order mode loss. In this context, for the convenience of the block diagram representation, each vacuum field is introduced as a two-photon quadrature defined at the position of the front mirror, incorporating the phase accumulation during spatial propagation from the original single-photon fields. It should be noted that the vacuum field $q_{\rm EL}^{\rm diff}$, which enters from the right-hand side of the end mirror, is not shown in Figure \ref{fig.4}. This is because the final detection port is located on the left-hand side of the input mirror, so this vacuum field does not contribute to the sensitivity.
In addition, since the function given in Equation \ref{eq.3-2} is implicitly defined on the right-hand side of the input mirror, amplitude and phase corrections must be applied in order to treat different quantum fields entering the cavity as if they are injected from the same spatial point. These corrections are implemented via compensation blocks placed before quantum fluctuations associated with loss are coupled into the system. For example, for the vacuum field $q_{\rm EL}^{\rm diff}$, the following compensation function $\mathcal{K}(\omega)$ is applied:
\begin{equation}
\label{eq.3-7}
{\cal K}(\omega) = U_{\rm{E}}D_{\rm{E}}\times\left[{r_{\rm{E}}({\omega})D_{\rm{E}}^2\exp\left({\rm{i}}\frac{L\omega}{c}+{\rm{i}} \frac{d\phi}{2}\right)}\right]^{-1}\period
\end{equation}

\begin{comment}
\begin{table}[t]
  \centering
  \caption{Parameters used in the simulation.}
  \label{table:sim_params}
  \begin{tabular}{lccc}
    \toprule
    Meaning           & Symbol        & \multicolumn{1}{c}{Value} & Units \\
    \midrule
    Laser Power       & $I$           & $10$   & W\\
    Laser Wavelength  & $\lambda$     & $515$  & nm\\
    Mirror Mass       & $M_{\mathrm{in}},\ M_{\mathrm{end}}$  & $100$  & kg\\
    Mirror Radius     & $R$           & $0.5$  & m\\
    Cavity Length     & $L$           & $1000$ & km \\
    Finesse           & $\mathcal{F}$ & $10$   & \\
    Loss Factor       & $D_{\rm{I}}, D_{\rm{E}}$ & $0.9760$ & \\
    \bottomrule
  \end{tabular}
\end{table}
\end{comment}

Furthermore, in the block diagram, $M$ denotes the mirror mass, and $A_i$ represents the amplitude of the optical field incident on the mirror. In other words, the aforementioned amplitude quantum fluctuations are converted into a force dimension through $G_{11}$, by a factor of $2{\hbar}{\omega}_0 A_i / c$. This force is then converted into displacement via $1 / (M_i s^2)$. Since gravitational wave signals manifest as mirror displacements in the interferometer, they are introduced into the block diagram at this stage. These displacements, in turn, couple to phase fluctuations by a factor of $2 A_ik_0$. Because the final gravitational wave signal appears in the phase quadrature, we calculate the contribution of each fluctuation to this component. In addition, following the standard DECIGO design, we assume a laser power of $10$ W, a wavelength of $515$ nm, and a cavity length of $1000$ km. Both mirrors forming the cavity are assumed to have identical geometry, with a mass of $100$ kg and a radius of $0.5$ m. The reflectivity is adjusted such that the cavity finesse is set to $10$. Given the assumption of identical mirror shapes, the beam waist is located at the midpoint between the two mirrors, that is, at a distance of $L/2$ from the input mirror. In this configuration, the parameter $D$, which represents the fraction of laser power remaining in the cavity after accounting for diffraction effects, can be analytically maximized. Following previous studies, we adopt this maximized value of $D = 0.9760$ as the baseline design in this paper.\par
Based on the standard design of DECIGO, the quantum noise spectrum with a detuning angle of $d\phi = 0$ rad is shown in Figure \ref{fig.5}. The shape of the curve is similar to that obtained without considering diffraction-related loss. It consists of radiation pressure noise, which follows an $f^{-2}$ frequency dependence and originates from amplitude quadrature fluctuations, and shot noise, which shows an $f^{0}$ (flat) dependence and arises from phase quadrature fluctuations. However, the contribution of each fluctuation component to the shot noise depends on where it enters the system, as the resulting phase shifts differ. Consequently, the corner frequencies vary among the components. A similar variation occurs for the amplitude quadrature fluctuations contributing to radiation pressure noise. However, the corresponding corner frequencies lie outside the range shown in the figure.
%%=%-%=%=%=%=%=%=%=%=%=%=%=%-%=%=%=%=%=%=%=%=%=%=%=%-%=%=%=%=%=%=%=%=%=%=%=%-%=%=%=%=%=%=%=%=%=%=%=%-%=%=%=%=%=%=%=%=%=%=%=%-%=%=%=%=%=%=%=%=%=%=%
%%=%-%=%=%=%=%=%=%=%=%=%=%=%-%=%=%=%=%=%=%=%=%=%=%=%-%=%=%=%=%=%=%=%=%=%=%=%-%=%=%=%=%=%=%=%=%=%=%=%-%=%=%=%=%=%=%=%=%=%=%=%-%=%=%=%=%=%=%=%=%=%=%
%%=%-%=%=%=%=%=%=%=%=%=%=%=%-%=%=%=%=%=%=%=%=%=%=%=%-%=%=%=%=%=%=%=%=%=%=%=%-%=%=%=%=%=%=%=%=%=%=%=%-%=%=%=%=%=%=%=%=%=%=%=%-%=%=%=%=%=%=%=%=%=%=%

\begin{figure*}[!t]
    \begin{center}
        \includegraphics[height=85mm]{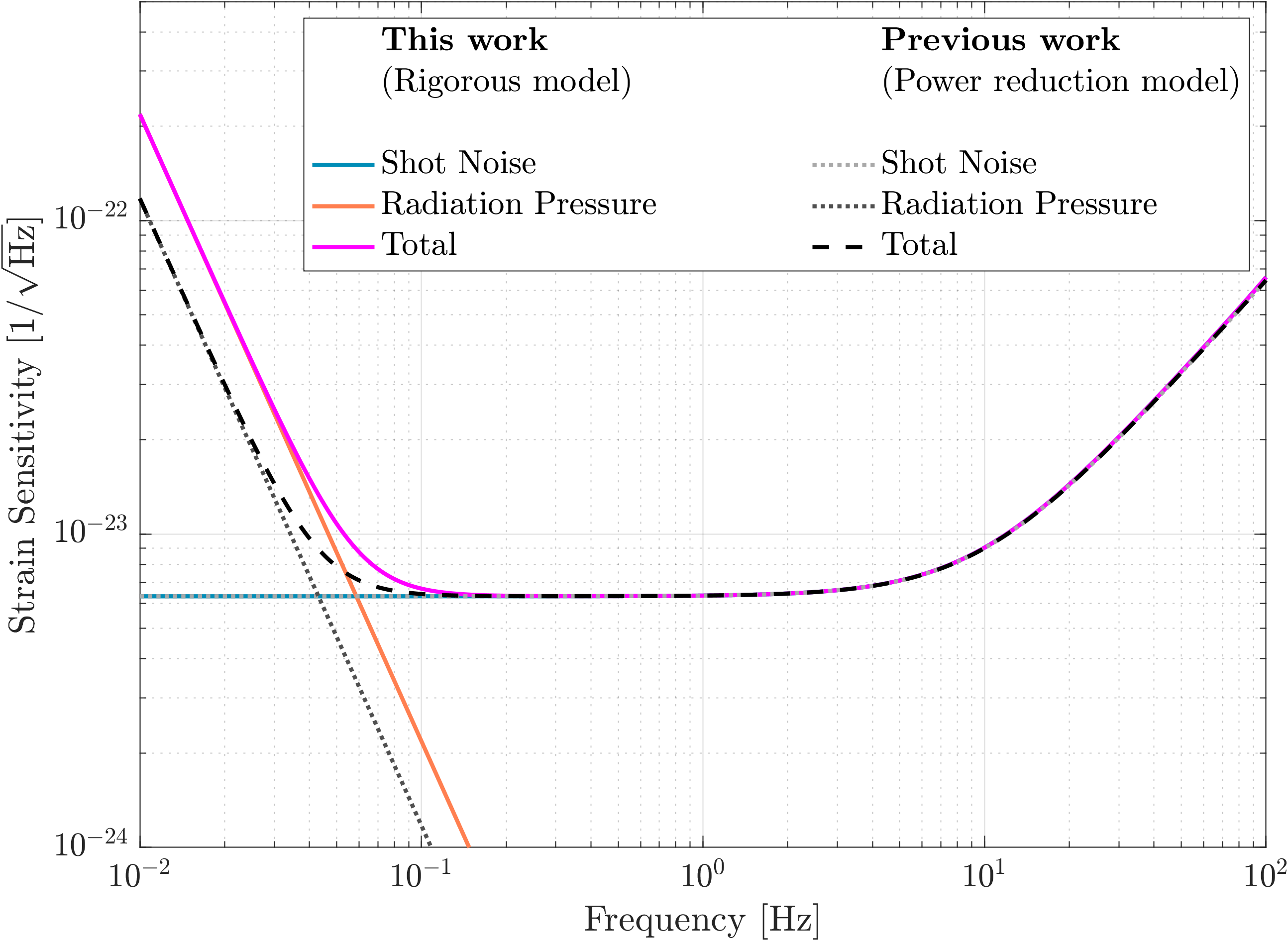}
        \caption{Comparison of the spectra of quantum noise in terms of strain sensitivity between the results obtained in this paper and those from previous research. The magenta solid curve represents the total quantum noise obtained using the rigorous formula derived in this paper, while the blue solid curve represents shot noise, and the orange solid curve represents radiation pressure noise obtained in the same manner. The black dotted curve represents the total quantum noise obtained using the power reduction model adopted in the previous study, while the light gray dotted curve represents shot noise, and the dark gray dotted curve represents radiation pressure noise in that previous model.}
        \label{fig.previous}
    \end{center}
\end{figure*}
\begin{figure*}[!t]
  \centering
  \hspace{-2mm}
  \begin{subfigure}[b]{.49\textwidth}
    \centering
    \includegraphics[width=\textwidth]{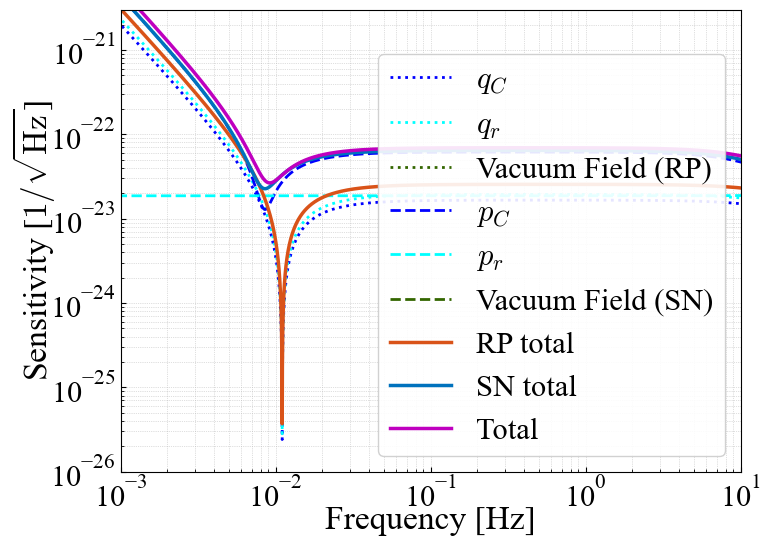}
    \caption{$D=1$, $d\phi=0.93369$}
    \label{fig:7a}
  \end{subfigure}
  \hspace{0mm}
  \begin{subfigure}[b]{.49\textwidth}
    \centering
    \includegraphics[width=\textwidth]{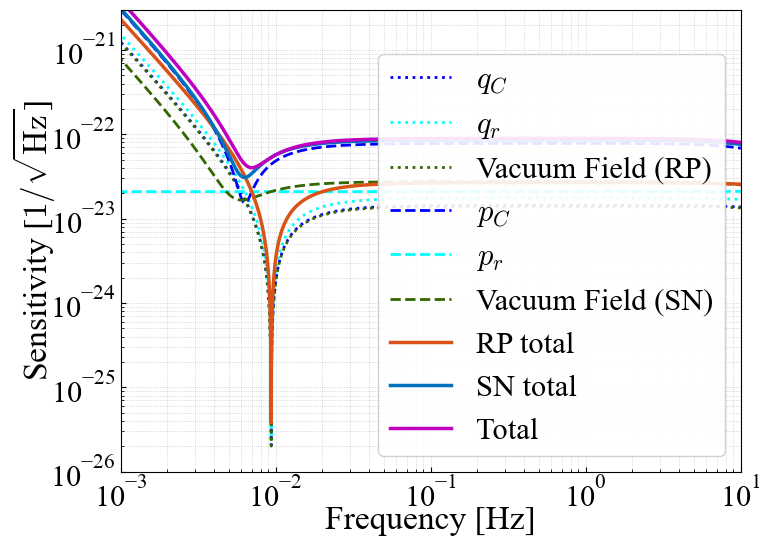}
    \caption{$D=0.97604$, $d\phi=1.1701$}
    \label{fig:7b}
  \end{subfigure}

  \vspace{1em}

  \begin{subfigure}[b]{.49\textwidth}
    \centering
    \includegraphics[width=\textwidth]{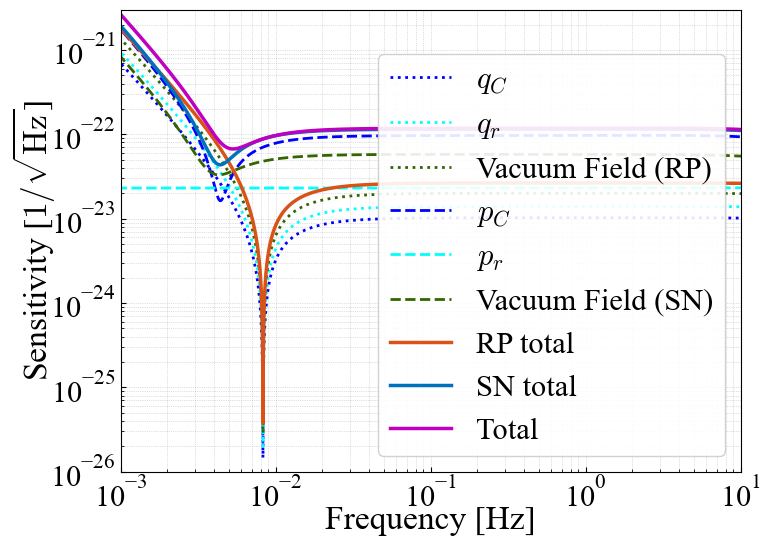}
    \caption{$D=0.92629$, $d\phi=1.2958$}
    \label{fig:7c}
  \end{subfigure}
  \hspace{0mm}
  \begin{subfigure}[b]{.49\textwidth}
    \centering
    \includegraphics[width=\textwidth]{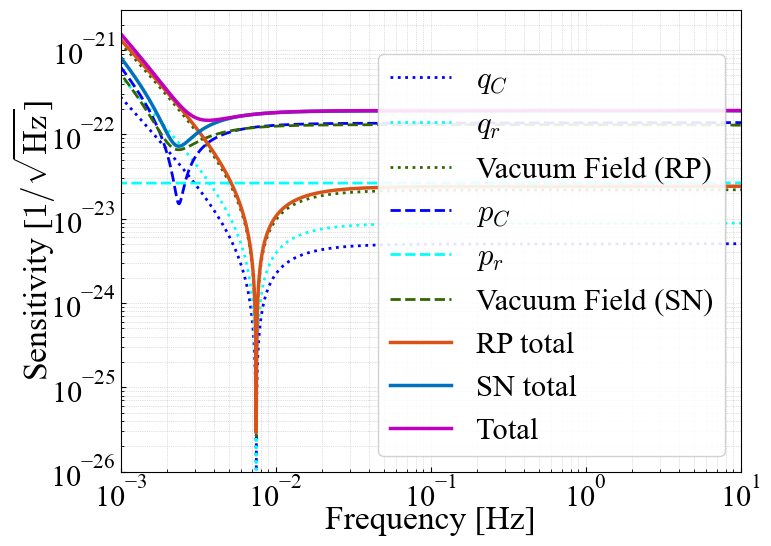}
    \caption{$D=0.81638$, $d\phi=1.4847$}
    \label{fig:7d}
  \end{subfigure}

  \caption{Spectra of quantum noise in terms of strain sensitivity in the presence of diffraction and higher-order mode loss, under optimized detuning angles of the main cavity. The top-left panel shows the case without diffraction effects, while the subsequent panels moving toward the bottom-right show the spectra of the quantum noise as diffraction and higher-order mode losses progressively increase. In each panel, dotted curves represent noise arising from various kinds of quantum fluctuations in the amplitude quadrature, with their total quadratic sum indicated by the orange solid curve (labeled “RP total”). Similarly, dashed curves represent noise from various kinds of quantum fluctuations in the phase quadrature, with their total quadratic sum shown as the blue solid curve (“SN total”). The resulting total quantum noise curve is depicted in purple (“Total”). Each panel also indicates diffraction factor $D$ and detuning angle $d\phi$.}
\label{fig.comparison}
\end{figure*}

\section{Discussion}\label{sec:discussion}
As shown in Figure~\ref{fig.previous}, we compare the quantum noise in terms of strain sensitivity obtained using the rigorous formulation presented in Section~\ref{sec:3} with the quantum noise evaluated in previous work \cite{galaxies9010009}. The previous work calculated shot noise and radiation pressure noise simply from the reduced laser power caused by diffraction effects, with the assumption that all relevant quantum fluctuations originate solely from this effective laser power. In contrast, the present study incorporates additional quantum fluctuations introduced via vacuum fields coupled through diffraction-related loss, including both diffraction loss and higher-order mode loss in the evaluation of both shot noise and radiation pressure noise.\par
As illustrated in the figure, the quantum noise derived in this study shows a slight degradation compared to previous results. This is primarily due to an increase in radiation pressure noise. The discrepancy arises because previous calculations considered only laser fluctuations as the driving force acting on the mirrors, thereby neglecting the contributions from other vacuum fields introduced by diffraction and higher-order mode losses. As a result, radiation pressure noise was underestimated in those works. In contrast, the present analysis includes these vacuum fields as additional sources of radiation pressure, leading to a more accurate and comprehensive representation of their impact.\par
In contrast to the increase observed in radiation pressure noise, no significant difference is observed in the shot noise between the previous and present analysis. This is because the phase quadrature fluctuations determining shot noise were effectively preserved even in the previous analysis. Consequently, the shot noise remains essentially unchanged and is generally robust against changes in the readout port, provided the detected quadrature and optical gain remain equivalent.\par
As a result, the overall sensitivity deteriorates slightly in the frequency range below $0.1$ Hz, where radiation pressure noise dominates. However, no significant change is observed in the range from $0.1$ Hz to $1$ Hz, which is DECIGO’s main target band. Therefore, we consider that this effect does not pose a serious concern for the baseline design of DECIGO. Additionally, since DECIGO is basically composed of simple Fabry-Perot cavities, the quantum noise power spectra obtained for individual cavities that do not consider effects such as mirror scattering can be used to evaluate the overall DECIGO sensitivity, following the method presented for example in T. Ishikawa et al. \cite{galaxies9010014}.\par

\begin{comment}
\begin{figure}[t]
    \centering
    \includegraphics[width=1\linewidth]{fig/fig8.JPG}
    \caption{Spectra of quantum noise in terms of strain sensitivity in the presence of diffraction and higher-order mode loss, under optimized detuning angles of the main cavity. The top-left panel shows the case without diffraction effects, while the subsequent panels moving toward the bottom-right show the spectra of the quantum noise as diffraction and higher-order mode losses progressively increase. In each panel, dotted curves represent noise arising from various kinds of quantum fluctuations in the amplitude quadrature, with their total quadratic sum indicated by the orange solid curve (labeled “RP total”). Similarly, dashed curves represent noise from various kinds of quantum fluctuations in the phase quadrature, with their total quadratic sum shown as the blue solid curve (“SN total”). The resulting total quantum noise curve is depicted in purple (“Total”). Each panel also indicates diffraction factor $D$ and detuning angle $d\phi$.}
    \label{fig.comparison}
\end{figure}
\end{comment}

Finally, we investigate the quantum noise when the cavity is detuned, while varying the contribution of vacuum fields associated with diffraction loss and higher-order mode loss. These variations correspond to changes in the beam size or the mirror size, assuming a constant mirror mass. Figure~\ref{fig.comparison} presents four representative quantum noise spectrum in terms of strain sensitivity obtained under different conditions on diffraction factor $D$ and detuning angle $d\phi$.\par
In all cases, dips appear in the quantum noise spectrum due to the detuning effect. However, the dip frequencies associated with each kind of quantum fluctuation do not completely coincide. This occurs because quantum noise sources for radiation pressure noise and shot noise enter the cavity at different spatial locations. These entry points correspond to distinct positions in the block diagram (Figure~\ref{fig.4}), resulting in different frequency dependencies. For instance, the dip caused by the phase fluctuation of the laser light originates from the reflected light at the input mirror and is independent of other fluctuations.\par
Meanwhile, the fact that dips also appear in the quantum noise spectrum associated with vacuum field fluctuations suggests that increasing the number of tunable optical parameters of the cavity could allow alignment of the dip frequencies of radiation pressure noise and shot noise. This may lead to further sensitivity optimization, even in the presence of diffraction-related loss. An advantage of this approach is that it does not require additional compensatory optics, such as those used in quantum locking techniques. Moreover, combining this method with quantum locking could enhance the feasibility of detecting primordial gravitational waves.\par
In summary, this study shows that while additional vacuum fields from diffraction-related loss increase radiation pressure noise, shot noise remains largely unchanged. Although a slight sensitivity degradation is seen at low frequencies, the impact on DECIGO’s main target band is negligible. Furthermore, the presence of dips in the detuned quantum noise spectrum suggests that sensitivity could be further optimized by adjusting optical parameters. This theoretical framework can be experimentally tested in ground-based interferometers like Advanced LIGO by scaling the beam size, providing a practical way to verify the model despite low-frequency technical noise.

%%=%-%=%=%=%=%=%=%=%=%=%=%=%-%=%=%=%=%=%=%=%=%=%=%=%-%=%=%=%=%=%=%=%=%=%=%=%-%=%=%=%=%=%=%=%=%=%=%=%-%=%=%=%=%=%=%=%=%=%=%=%-%=%=%=%=%=%=%=%=%=%=%
%%=%-%=%=%=%=%=%=%=%=%=%=%=%-%=%=%=%=%=%=%=%=%=%=%=%-%=%=%=%=%=%=%=%=%=%=%=%-%=%=%=%=%=%=%=%=%=%=%=%-%=%=%=%=%=%=%=%=%=%=%=%-%=%=%=%=%=%=%=%=%=%=%
%%=%-%=%=%=%=%=%=%=%=%=%=%=%-%=%=%=%=%=%=%=%=%=%=%=%-%=%=%=%=%=%=%=%=%=%=%=%-%=%=%=%=%=%=%=%=%=%=%=%-%=%=%=%=%=%=%=%=%=%=%=%-%=%=%=%=%=%=%=%=%=%=%
\section{Conclusion}\label{sec:summary}

Precise characterization of quantum field modifications induced by diffraction effects is essential for evaluating the sensitivity of interferometric detectors, particularly in long-baseline cavities such as those adopted in DECIGO. In this paper, the influence of diffraction effects was formulated by introducing vacuum fields that are independent of the intrinsic quantum fluctuations of the laser field. This advancement is particularly important for next-generation detectors, where long-baseline cavities make diffraction effects non-negligible.\par
Based on this formalism, the sensitivity was calculated and compared with a conventional approach in which diffraction effects are treated as a simple reduction in optical power. The resulting quantum noise showed that, although the shot noise level remained unchanged, the radiation pressure noise was enhanced. We found that this enhancement originated from the amplitude fluctuations of the additional vacuum fields, which exert an extra quantum back-action on the mirrors. As a result, this behavior is a distinctive feature of the present framework, in contrast to the conventional treatment of diffraction effects. Unlike conventional approaches, our framework naturally accounts for all quantum fluctuations induced by optical losses without relying on empirical correction factors.\par
It was also confirmed that, even under this rigorous treatment, detuning the cavity enables the quantum noise spectrum to exhibit a dip, similar to what is observed in cavities without diffraction effects. However, the dip is rather mild, due to the overlap of various quantum noise components that have different dip frequencies. This result indicates that, as in previous studies, parameter optimization is likely to play an important role in improving sensitivity. \par
Importantly, by establishing a precise treatment of quantum noise in the presence of diffraction, this work enables a detailed investigation of how much the sensitivity can be enhanced in DECIGO by detuning the main cavity and employing homodyne detection without the need for auxiliary cavities, and whether implementing such a configuration will improve the chances of detecting PGWs \cite{tsuji2025quantumnoisereductionspacebased}. Furthermore, this framework also allows us to explore whether combining the detuning of the main cavity with homodyne detection in optical-spring quantum locking systems that use auxiliary cavities can provide additional sensitivity improvement, thereby enhancing the feasibility of detecting primordial gravitational waves. Therefore, it lays a critical foundation for unlocking the full potential of future space-based detectors such as DECIGO.

%%=%-%=%=%=%=%=%=%=%=%=%=%=%-%=%=%=%=%=%=%=%=%=%=%=%-%=%=%=%=%=%=%=%=%=%=%=%-%=%=%=%=%=%=%=%=%=%=%=%-%=%=%=%=%=%=%=%=%=%=%=%-%=%=%=%=%=%=%=%=%=%=%
%%=%-%=%=%=%=%=%=%=%=%=%=%=%-%=%=%=%=%=%=%=%=%=%=%=%-%=%=%=%=%=%=%=%=%=%=%=%-%=%=%=%=%=%=%=%=%=%=%=%-%=%=%=%=%=%=%=%=%=%=%=%-%=%=%=%=%=%=%=%=%=%=%
%%=%-%=%=%=%=%=%=%=%=%=%=%=%-%=%=%=%=%=%=%=%=%=%=%=%-%=%=%=%=%=%=%=%=%=%=%=%-%=%=%=%=%=%=%=%=%=%=%=%-%=%=%=%=%=%=%=%=%=%=%=%-%=%=%=%=%=%=%=%=%=%=%

%%=%-%=%=%=%=%=%=%=%=%=%=%=%-%=%=%=%=%=%=%=%=%=%=%=%-%=%=%=%=%=%=%=%=%=%=%=%-%=%=%=%=%=%=%=%=%=%=%=%-%=%=%=%=%=%=%=%=%=%=%=%-%=%=%=%=%=%=%=%=%=%=%
%%=%-%=%=%=%=%=%=%=%=%=%=%=%-%=%=%=%=%=%=%=%=%=%=%=%-%=%=%=%=%=%=%=%=%=%=%=%-%=%=%=%=%=%=%=%=%=%=%=%-%=%=%=%=%=%=%=%=%=%=%=%-%=%=%=%=%=%=%=%=%=%=%
%%=%-%=%=%=%=%=%=%=%=%=%=%=%-%=%=%=%=%=%=%=%=%=%=%=%-%=%=%=%=%=%=%=%=%=%=%=%-%=%=%=%=%=%=%=%=%=%=%=%-%=%=%=%=%=%=%=%=%=%=%=%-%=%=%=%=%=%=%=%=%=%=%
\vspace{1mm}
\noindent\textbf{Funding}\\
This work was supported by JSPS KAKENHI, Grant No. JP22H01247/23K22518 and 25KJ1390.
\vspace{1mm}

\noindent\textbf{Acknowledgment}\\
 We would like to thank Nikitha Kuntimaddi for English editing.
\vspace{1mm}

\noindent\textbf{Disclosures}\\
The authors declare no conflicts of interest.
\vspace{1mm}

\noindent\textbf{Data availability}\\
Data underlying the results presented in this paper are not publicly available at this time but may be obtained from the authors upon reasonable request.

%%%%%%%%%%%%%%%%%%%%%%% References %%%%%%%%%%%%%%%%%%%%%%%%%

%%%%%%%%%% If using BibTeX:
\clearpage
\bibliography{REF_LIST}

@Article{10.1093/ptep/ptab019,
  author   = {Kawamura, Seiji and Ando, Masaki and Seto, Naoki and Sato, Shuichi and Musha, Mitsuru and Kawano, Isao and Yokoyama, Jun'ichi and Tanaka, Takahiro and Ioka, Kunihito and Akutsu, Tomotada and others},
  journal  = {Progress of Theoretical and Experimental Physics},
  title    = {{Current status of space gravitational wave antenna DECIGO and B-DECIGO}},
  year     = {2021},
  issn     = {2050-3911},
  month    = {02},
  note     = {05A105},
  number   = {5},
  volume   = {2021},
  doi      = {10.1093/ptep/ptab019},
  eprint   = {https://academic.oup.com/ptep/article-pdf/2021/5/05A105/38109685/ptab019.pdf},
  url      = {https://doi.org/10.1093/ptep/ptab019},
}

@Article{PhysRevLett.87.221103,
  author    = {Seto, Naoki and Kawamura, Seiji and Nakamura, Takashi},
  journal   = {Phys. Rev. Lett.},
  title     = {Possibility of Direct Measurement of the Acceleration of the Universe Using 0.1 Hz Band Laser Interferometer Gravitational Wave Antenna in Space},
  year      = {2001},
  month     = {Nov},
  pages     = {221103},
  volume    = {87},
  doi       = {10.1103/PhysRevLett.87.221103},
  issue     = {22},
  numpages  = {4},
  publisher = {American Physical Society},
  url       = {https://link.aps.org/doi/10.1103/PhysRevLett.87.221103},
}

@Article{galaxies12020013,
  author         = {Tsuji, Kenji and Ishikawa, Tomohiro and Umemura, Kurumi and Kawasaki, Yuki and Iwaguchi, Shoki and Shimizu, Ryuma and Ando, Masaki and Kawamura, Seiji},
  journal        = {Galaxies},
  title          = {Significance of Fabry-Perot Cavities for Space Gravitational Wave Antenna DECIGO},
  year           = {2024},
  issn           = {2075-4434},
  number         = {2},
  volume         = {12},
  article-number = {13},
  doi            = {10.3390/galaxies12020013},
  url            = {https://www.mdpi.com/2075-4434/12/2/13},
}

@Article{akrami2020planck,
  author  = {Akrami Cheghasiahi, Yashar and Arroja, F and Ashdown, M and Aumont, J and Baccigalupi, Carlo and Ballardini, M and Banday, Anthony J and Barreiro, RB and Bartolo, N and Basak, S and others},
  journal = {Astronomy and Astrophysics (A \& A)},
  title   = {Planck 2018 results: X. Constraints on inflation},
  year    = {2020},
  volume  = {641},
}

@Article{YAMADA2020126626,
  author   = {Rika Yamada and Yutaro Enomoto and Atsushi Nishizawa and Koji Nagano and Sachiko Kuroyanagi and Keiko Kokeyama and Kentaro Komori and Yuta Michimura and Takeo Naito and Izumi Watanabe and Taigen Morimoto and Masaki Ando and Akira Furusawa and Seiji Kawamura},
  journal  = {Physics Letters A},
  title    = {Optimization of quantum noise by completing the square of multiple interferometer outputs in quantum locking for gravitational wave detectors},
  year     = {2020},
  issn     = {0375-9601},
  number   = {26},
  pages    = {126626},
  volume   = {384},
  doi      = {https://doi.org/10.1016/j.physleta.2020.126626},
  url      = {https://www.sciencedirect.com/science/article/pii/S037596012030493X},
}

@Article{YAMADA2021127365,
  author   = {Rika Yamada and Yutaro Enomoto and Izumi Watanabe and Koji Nagano and Yuta Michimura and Atsushi Nishizawa and Kentaro Komori and Takeo Naito and Taigen Morimoto and Shoki Iwaguchi and Tomohiro Ishikawa and Masaki Ando and Akira Furusawa and Seiji Kawamura},
  journal  = {Physics Letters A},
  title    = {Reduction of quantum noise using the quantum locking with an optical spring for gravitational wave detectors},
  year     = {2021},
  issn     = {0375-9601},
  pages    = {127365},
  volume   = {402},
  doi      = {https://doi.org/10.1016/j.physleta.2021.127365},
  url      = {https://www.sciencedirect.com/science/article/pii/S0375960121002292},
}

@Article{galaxies9010009,
  author         = {Iwaguchi, Shoki and Ishikawa, Tomohiro and Ando, Masaki and Michimura, Yuta and Komori, Kentaro and Nagano, Koji and Akutsu, Tomotada and Musha, Mitsuru and Yamada, Rika and Watanabe, Izumi and others},
  journal        = {Galaxies},
  title          = {Quantum Noise in a Fabry-Perot Interferometer Including the Influence of Diffraction Loss of Light},
  year           = {2021},
  issn           = {2075-4434},
  number         = {1},
  volume         = {9},
  article-number = {9},
  doi            = {10.3390/galaxies9010009},
  url            = {https://www.mdpi.com/2075-4434/9/1/9},
}

@Article{galaxies9010014,
  author         = {Ishikawa, Tomohiro and Iwaguchi, Shoki and Michimura, Yuta and Ando, Masaki and Yamada, Rika and Watanabe, Izumi and Nagano, Koji and Akutsu, Tomotada and Komori, Kentaro and Musha, Mitsuru and others},
  journal        = {Galaxies},
  title          = {Improvement of the Target Sensitivity in DECIGO by Optimizing Its Parameters for Quantum Noise Including the Effect of Diffraction Loss},
  year           = {2021},
  issn           = {2075-4434},
  number         = {1},
  volume         = {9},
  article-number = {14},
  doi            = {10.3390/galaxies9010014},
  url            = {https://www.mdpi.com/2075-4434/9/1/14},
}

@Article{galaxies10010025,
  author         = {Kawasaki, Y. and Shimizu, R. and Ishikawa, T. and Nagano, K. and Iwaguchi, S. and others},
  journal        = {Galaxies},
  title          = {Optimization of Design Parameters for Gravitational Wave Detector {DECIGO} Including Fundamental Noises},
  year           = {2022},
  issn           = {2075-4434},
  number         = {1},
  volume         = {10},
  article-number = {25},
  doi            = {10.3390/galaxies10010025},
  url            = {https://www.mdpi.com/2075-4434/10/1/25},
}

@Article{galaxies11060111,
  author         = {Tsuji, Kenji and Ishikawa, Tomohiro and Komori, Kentaro and Nagano, Koji and Enomoto, Yutaro and Michimura, Yuta and Umemura, Kurumi and Shimizu, Ryuma and Wu, Bin and Iwaguchi, Shoki and others},
  journal        = {Galaxies},
  title          = {Optimization of Quantum Noise in Space Gravitational-Wave Antenna DECIGO with Optical-Spring Quantum Locking Considering Mixture of Vacuum Fluctuations in Homodyne Detection},
  year           = {2023},
  issn           = {2075-4434},
  number         = {6},
  volume         = {11},
  article-number = {111},
  doi            = {10.3390/galaxies11060111},
  url            = {https://www.mdpi.com/2075-4434/11/6/111},
}

@Book{schleich2011quantum,
  author    = {Schleich, Wolfgang P},
  publisher = {John Wiley \& Sons},
  title     = {Quantum optics in phase space},
  year      = {2011},
}

@Article{PhysRevLett.90.083601,
  author    = {Courty, J. and Heidmann, A. and Pinard, M.},
  journal   = {Phys. Rev. Lett.},
  title     = {Quantum Locking of Mirrors in Interferometers},
  year      = {2003},
  month     = {Feb},
  pages     = {083601},
  volume    = {90},
  doi       = {10.1103/PhysRevLett.90.083601},
  issue     = {8},
  numpages  = {4},
  publisher = {American Physical Society},
  url       = {https://link.aps.org/doi/10.1103/PhysRevLett.90.083601},
}

@Article{Antoine_Heidmann_2004,
  author   = {Heidmann, A. and Courty, J. and Pinard, M. and Lebars, J.},
  journal  = {Journal of Optics B: Quantum and Semiclassical Optics},
  title    = {Beating quantum limits in interferometers with quantum locking of mirrors},
  year     = {2004},
  month    = {Jul},
  number   = {8},
  pages    = {S684},
  volume   = {6},
  doi      = {10.1088/1464-4266/6/8/009},
  url      = {https://dx.doi.org/10.1088/1464-4266/6/8/009},
}

@Article{PhysRevD.107.022007,
  author    = {Ishikawa, T. and Kawasaki, Y. and Tsuji, K. and Yamada, R. and Watanabe, I. and others},
  journal   = {Phys. Rev. D},
  title     = {First-step experiment for sensitivity improvement of {DECIGO}: Sensitivity optimization for simulated quantum noise by completing the square},
  year      = {2023},
  month     = {Jan},
  pages     = {022007},
  volume    = {107},
  doi       = {10.1103/PhysRevD.107.022007},
  issue     = {2},
  numpages  = {9},
  publisher = {American Physical Society},
  url       = {https://link.aps.org/doi/10.1103/PhysRevD.107.022007},
}

@Article{Ishikawa_2024,
  author    = {Ishikawa, T and Kawasaki, Y and Tsuji, K and Shimizu, R and Umemura, K and Wu, B and Iwaguchi, S and Michimura, Y and Nagano, K and Enomoto, Y and Komori, K and Doki, S and Furusawa, A and Kawamura, S},
  journal   = {Classical and Quantum Gravity},
  title     = {Feasibility of loop-gain tuning for general measurement systems inspired by quantum locking for DECIGO},
  year      = {2024},
  month     = {oct},
  number    = {21},
  pages     = {215013},
  volume    = {41},
  doi       = {10.1088/1361-6382/ad7cb6},
  publisher = {IOP Publishing},
  url       = {https://dx.doi.org/10.1088/1361-6382/ad7cb6},
}

@Article{PhysRevA.57.2134,
  author    = {Barnett, Stephen M. and Jeffers, John and Gatti, Alessandra and Loudon, Rodney},
  journal   = {Phys. Rev. A},
  title     = {Quantum optics of lossy beam splitters},
  year      = {1998},
  month     = {Mar},
  pages     = {2134--2145},
  volume    = {57},
  doi       = {10.1103/PhysRevA.57.2134},
  issue     = {3},
  numpages  = {0},
  publisher = {American Physical Society},
  url       = {https://link.aps.org/doi/10.1103/PhysRevA.57.2134},
}

@Article{PhysRevLett.77.1739,
  author    = {Barnett, Stephen M. and Gilson, Claire R. and Huttner, Bruno and Imoto, Nobuyuki},
  journal   = {Phys. Rev. Lett.},
  title     = {Field Commutation Relations in Optical Cavities},
  year      = {1996},
  month     = {Aug},
  pages     = {1739--1742},
  volume    = {77},
  doi       = {10.1103/PhysRevLett.77.1739},
  issue     = {9},
  numpages  = {0},
  publisher = {American Physical Society},
  url       = {https://link.aps.org/doi/10.1103/PhysRevLett.77.1739},
}

@Article{tsuji2025quantumnoisereductionspacebased,
      title         = {Quantum Noise Reduction in the Space-based Gravitational Wave Antenna DECIGO Using Optical Springs and Homodyne Detection scheme}, 
      author        = {Kenji Tsuji and Tomohiro Ishikawa and Kentaro Komori and Yutaro Enomoto and Yuta Michimura and Kurumi Umemura and Shoki Iwaguchi and Keiko Kokeyama and Seiji Kawamura},
      year          = {2025},
      eprint        = {2509.17372},
      journal = {arXiv},
      primaryClass  = {gr-qc},
      url           = {https://arxiv.org/abs/2509.17372}, 
}

@article{danilishin2012quantum,
  title={Quantum measurement theory in gravitational-wave detectors},
  author={Danilishin, Stefan L and Khalili, Farid Ya},
  journal={Living Reviews in Relativity},
  volume={15},
  number={1},
  pages={5},
  year={2012},
  publisher={Springer}
}

@article{PhysRevA.31.3068,
  title = {New formalism for two-photon quantum optics. I. Quadrature phases and squeezed states},
  author = {Caves, Carlton M. and Schumaker, Bonny L.},
  journal = {Phys. Rev. A},
  volume = {31},
  issue = {5},
  pages = {3068--3092},
  numpages = {0},
  year = {1985},
  month = {May},
  publisher = {American Physical Society},
  doi = {10.1103/PhysRevA.31.3068},
  url = {https://link.aps.org/doi/10.1103/PhysRevA.31.3068}
}

@article{PhysRevD.65.022002,
  title = {Conversion of conventional gravitational-wave interferometers into quantum nondemolition interferometers by modifying their input and/or output optics},
  author = {Kimble, H. J. and Levin, Yuri and Matsko, Andrey B. and Thorne, Kip S. and Vyatchanin, Sergey P.},
  journal = {Phys. Rev. D},
  volume = {65},
  issue = {2},
  pages = {022002},
  numpages = {31},
  year = {2001},
  month = {Dec},
  publisher = {American Physical Society},
  doi = {10.1103/PhysRevD.65.022002},
  url = {https://link.aps.org/doi/10.1103/PhysRevD.65.022002}
}

@article{PhysRevD.64.042006,
  title = {Quantum noise in second generation, signal-recycled laser interferometric gravitational-wave detectors},
  author = {Buonanno, Alessandra and Chen, Yanbei},
  journal = {Phys. Rev. D},
  volume = {64},
  issue = {4},
  pages = {042006},
  numpages = {21},
  year = {2001},
  month = {Jul},
  publisher = {American Physical Society},
  doi = {10.1103/PhysRevD.64.042006},
  url = {https://link.aps.org/doi/10.1103/PhysRevD.64.042006}
}

@article{PhysRevA.72.013818,
  title = {Mathematical framework for simulation of quantum fields in complex interferometers using the two-photon formalism},
  author = {Corbitt, Thomas and Chen, Yanbei and Mavalvala, Nergis},
  journal = {Phys. Rev. A},
  volume = {72},
  issue = {1},
  pages = {013818},
  numpages = {16},
  year = {2005},
  month = {Jul},
  publisher = {American Physical Society},
  doi = {10.1103/PhysRevA.72.013818},
  url = {https://link.aps.org/doi/10.1103/PhysRevA.72.013818}
}

@article{PhysRevD.93.082004,
  title = {Estimation of losses in a 300 m filter cavity and quantum noise reduction in the KAGRA gravitational-wave detector},
  author = {Capocasa, Eleonora and Barsuglia, Matteo and Degallaix, J\'er\^ome and Pinard, Laurent and Straniero, Nicolas and Schnabel, Roman and Somiya, Kentaro and Aso, Yoichi and Tatsumi, Daisuke and Flaminio, Raffaele},
  journal = {Phys. Rev. D},
  volume = {93},
  issue = {8},
  pages = {082004},
  numpages = {11},
  year = {2016},
  month = {Apr},
  publisher = {American Physical Society},
  doi = {10.1103/PhysRevD.93.082004},
  url = {https://link.aps.org/doi/10.1103/PhysRevD.93.082004}
}

@article{PhysRevD.88.022002,
  title = {Realistic filter cavities for advanced gravitational wave detectors},
  author = {Evans, M. and Barsotti, L. and Kwee, P. and Harms, J. and Miao, H.},
  journal = {Phys. Rev. D},
  volume = {88},
  issue = {2},
  pages = {022002},
  numpages = {7},
  year = {2013},
  month = {Jul},
  publisher = {American Physical Society},
  doi = {10.1103/PhysRevD.88.022002},
  url = {https://link.aps.org/doi/10.1103/PhysRevD.88.022002}
}
%\input{sa2.bbl}
%%%%%%%%%% If preparing manually:
% \begin{thebibliography}{1}
% \newcommand{\enquote}[1]{``#1''}

% \bibitem{Zhang:14}
% Y.~Zhang, S.~Qiao, L.~Sun, Q.~W. Shi, W.~Huang, L.~Li, and Z.~Yang,
%   \enquote{Photoinduced active terahertz metamaterials with nanostructured
%   vanadium dioxide film deposited by sol-gel method,}
%   {\protect\JournalTitle{Optics Express}} \textbf{22}, 11070--11078 (2014).

% \bibitem{Optica}
% {Optica}, \enquote{{Optica Publishing Group},}
%   \url{http://www.opg.optica.org}.

% \bibitem{FORSTER2007}
% P.~Forster, V.~Ramaswamy, P.~Artaxo, T.~Bernsten, R.~Betts, D.~Fahey,
%   J.~Haywood, J.~Lean, D.~Lowe, G.~Myhre, J.~Nganga, R.~Prinn, G.~Raga,
%   M.~Schulz, and R.~V. Dorland, \enquote{Changes in atmospheric consituents and
%   in radiative forcing,} in \enquote{Climate Change 2007: The Physical Science
%   Basis. Contribution of Working Group 1 to the Fourth Assesment Report of
%   Intergovernmental Panel on Climate Change,}  S.~Solomon, D.~Qin, M.~Manning,
%   Z.~Chen, M.~Marquis, K.~B. Averyt, M.~Tignor, and H.~L. Miler, eds.
%   (Cambridge University Press, 2007).

% \end{thebibliography}

\end{document}